\documentclass[10pt,journal,compsoc]{IEEEtran}

\usepackage{subcaption}
\usepackage{comment}
\usepackage{amsmath,amssymb} 
\usepackage{color}
\usepackage{dsfont}
\usepackage{booktabs}
\usepackage{multirow}
\usepackage{xcolor}
\usepackage{colortbl}
\usepackage{graphicx}
\usepackage{paralist}
\usepackage{marvosym}
\usepackage[export]{adjustbox}
\usepackage{makecell}

\newcommand{\ch}{\checkmark}
\definecolor{sh_gray}{rgb}{0.84,0.84,0.84}
\definecolor{sh_gray2}{rgb}{1,0.89,0.75}
\definecolor{color3}{rgb}{0.95,0.95,0.95}
\definecolor{color4}{rgb}{0.96,0.96,0.86}
\definecolor{color5}{rgb}{0.90,0.90,0.90}

\newcolumntype{M}[1]{>{\centering\arraybackslash}m{#1}}
\newcommand{\good}[1]{{\color{red}($#1\downarrow$)}}  
\def\xnet{MIRNet-v2~}

\makeatother
\usepackage[pagebackref=true,breaklinks=true,letterpaper=true,colorlinks,bookmarks=false,citecolor=blue,linkcolor=blue,urlcolor=blue]{hyperref}

\ifCLASSOPTIONcompsoc
  \usepackage[nocompress]{cite}
\else
  \usepackage{cite}
\fi

\begin{document}

\title{Learning Enriched Features for Fast Image Restoration and Enhancement}

\author{Syed Waqas Zamir,
Aditya Arora,
Salman Khan,
Munawar Hayat,\\
Fahad Shahbaz Khan,
Ming-Hsuan Yang,
and Ling Shao
\IEEEcompsocitemizethanks{\IEEEcompsocthanksitem S.W.~Zamir, and A.~Arora, are with Inception Institute of Artificial Intelligence, UAE. E-mail: waqas.zamir@inceptioniai.org
\IEEEcompsocthanksitem S.~khan and F.S.~Khan are with Mohammed Bin Zayed University of Artificial Intelligence, UAE.
\IEEEcompsocthanksitem M.~Hayat is with Monash Univeristy, Melbourne, Australia.
\IEEEcompsocthanksitem M.-H.~Yang is with University of California at Merced, and Google, USA.
\IEEEcompsocthanksitem L.Shao is with Terminus Group, China. }
}

\markboth{IEEE Transactions on Pattern Analysis and Machine Intelligence}%
{Shell \MakeLowercase{\textit{Zamir et al.}}: Bare Demo of IEEEtran.cls for Computer Society Journals}

\IEEEtitleabstractindextext{%
\begin{abstract}
Given a degraded input image, image restoration aims to recover the missing high-quality image content. 
Numerous applications demand effective image restoration, e.g., computational photography, surveillance, autonomous vehicles, and remote sensing.
Significant advances in image restoration have been made in recent years, dominated by convolutional neural networks (CNNs).
The widely-used CNN-based methods typically operate either on full-resolution or on progressively low-resolution representations. In the former case, spatial details are preserved but the contextual information cannot be precisely encoded.
In the latter case, generated outputs are semantically reliable but spatially less accurate. 
This paper presents a new architecture with a holistic goal of maintaining spatially-precise high-resolution representations through the entire network, and receiving complementary contextual information from the low-resolution representations. 
The core of our approach is a multi-scale residual block containing the following key elements: (a) parallel multi-resolution convolution streams for extracting multi-scale features, (b) information exchange across the multi-resolution streams, (c) non-local attention mechanism for capturing contextual information, and (d) attention based multi-scale feature aggregation. 
Our approach learns an enriched set of features that combines contextual information from multiple scales, while simultaneously preserving the high-resolution spatial details.
Extensive experiments on six real image benchmark datasets demonstrate that our method, named as \xnet, achieves state-of-the-art results for a variety of image processing tasks, including  defocus deblurring, image denoising, super-resolution, and image enhancement. The source code and pre-trained models are available at \url{https://github.com/swz30/MIRNetv2}.
\end{abstract}

}

\maketitle

\IEEEdisplaynontitleabstractindextext

%
\IEEEpeerreviewmaketitle

\IEEEraisesectionheading{\section{Introduction}\label{sec:introduction}}
\IEEEPARstart{O}wing to the physical limitations of cameras or due to complicated lighting conditions, image degradations of varying severity are often introduced as part of image acquisition.
For instance, smartphone cameras come with a narrow aperture and have small sensors with limited dynamic range. 
Consequently, they frequently generate noisy and low-contrast images. 
Similarly, images captured under the unsuitable lighting are either too dark or too bright.
Image restoration aims to recover the original clean image from its corrupted measurements.
It is an ill-posed inverse problem, due to the existence of many possible solutions.

Recent advances in image restoration and enhancement have been led by deep learning models, as they can learn strong (generalizable) priors from large-scale datasets. 
Existing CNNs typically follow one of the two architecture designs: 1) an encoder-decoder, or 2) high-resolution (single-scale) feature processing. 
The encoder-decoder models \cite{ronneberger2015u,kupyn2019deblurgan,chen2018,zhang2019kindling} first progressively map the input to a low-resolution representation, and then apply a gradual reverse mapping to the original resolution. 
Although these approaches learn a broad context by spatial-resolution reduction, on the downside, the fine spatial details are lost, making it extremely hard to recover them in the later stages. 
On the other hand, the high-resolution (single-scale) networks \cite{dong2015image,DnCNN,zhang2020residual,ignatov2017dslr} do not employ any downsampling operation, and thereby recover better spatial details. However, these networks have limited receptive field and are less effective in encoding contextual information.

\begin{table*}[t]
\begin{center}
\caption{Comparison between \xnet and MIRNet~\cite{zamir2020mirnet} under the same experimental settings for image denoising task on the SIDD benchmark dataset~\cite{sidd}. FLOPs and inference times are computed on an image of size $256$${\times}$$256$. When compared to MIRNet~\cite{zamir2020mirnet}, \xnet is more accurate, while being significantly lighter and faster. }
\label{table:compute comparisons}
\setlength{\tabcolsep}{8.5pt}
\scalebox{1}{
\begin{tabular}{l c l l l l l l }
\toprule[0.15em]
& PSNR & Params (M) & FLOPs (B) & Convs   & Activations (M)  & Train Time (h)   & Inference Time (ms) \\
\midrule[0.15em]
MIRNet~\cite{zamir2020mirnet} & 39.72 & 31.79 & 785 & 635 & 1270 & 139 & 142\\
\xnet (Ours) & 39.84 & 5.9 \good{81\%} & 140 \good{82\%} & 406 \good{36\%} & 390 \good{69\%} & 63 \good{55\%} & 39 \good{72\%}\\
\bottomrule[0.10em]
\end{tabular}}
\end{center}
\end{table*}

Image restoration is a position-sensitive procedure, where pixel-to-pixel correspondence from the input image to the output image is needed. 
Therefore, it is important to remove only the undesired degraded image content, while carefully preserving the desired fine spatial details (such as true edges and texture).
Such functionality for segregating the degraded content from the true signal can be better incorporated into CNNs with the help of large context, \textit{e.g.}, by enlarging the receptive field. 
Towards this goal, we develop a new \emph{multi-scale} approach that maintains the original high-resolution features along the network hierarchy, thus minimizing the loss of precise spatial details. Simultaneously, our model encodes multi-scale context by using \emph{parallel convolution streams} that process features at lower spatial resolutions. 
The multi-resolution parallel branches operate in a manner that is complementary to the main high-resolution branch, thereby providing us more precise and contextually enriched feature representations. 

One main distinction between our method and the existing multi-scale image processing approaches is how we aggregate contextual information. 
The existing methods \cite{tao2018scale,nah2017,gu2019self} process each scale in isolation. 
In contrast, we \textit{progressively} exchange and fuse information from coarse-to-fine resolution-levels. 
Furthermore, different from existing methods that employ a simple concatenation or averaging of features coming from multi-resolution branches, we introduce a new \emph{selective kernel} fusion approach that dynamically selects the useful set of kernels from each branch representations using a self-attention mechanism. 
More importantly, the proposed fusion block combines features with varying receptive fields, while preserving their distinctive complementary characteristics.

The main contributions of this work include:\vspace{-0.0em}
\begin{itemize}
\item A novel feature extraction model that obtains a complementary set of features across multiple spatial scales, while maintaining the original high-resolution features to preserve precise spatial details (Sec.~\ref{sec:method}).
\item A regularly repeated mechanism for information exchange, where the features from coarse-to-fine resolution branches are progressively fused together for improved representation learning (Sec.~\ref{sec:msrb}).
\item A new approach to fuse multi-scale features using a selective kernel network that dynamically combines variable receptive fields and faithfully preserves the original feature information at each spatial resolution (Sec.~\ref{sec:skff}).
\end{itemize}

 A preliminary version of this work has been published as a conference paper~\cite{zamir2020mirnet}. The MIRNet model~\cite{zamir2020mirnet} is expensive in terms of size and speed. In this work, we make several key modifications to MIRNet~\cite{zamir2020mirnet} that allow us to significantly reduce the computational cost while enhancing  model performance (see Table~\ref{table:compute comparisons}). Specifically, in the proposed \xnet, 
    \textbf{(a)} We demonstrate feature fusion only in the direction from low- to high-resolution streams performs best, and the information flow from high- to low-resolution branches can be removed to improve efficiency.
    \textbf{(b)} We replace the dual attention unit with a new residual contextual block (RCB). Furthermore, we introduce group convolutions in RCB that are capable of learning unique representations in each filter group, while being more resource efficient than standard convolutions. 
    \textbf{(c)} We employ progressive learning to improve training speed: the network is trained on small image patches in the early epochs and on gradually large patches in the later training epochs. 
    \textbf{(d)} We show the effectiveness of the proposed design on a new task of dual-pixel defocus deblurring~\cite{abdullah2020dpdd} {alongside the other image processing tasks of image denoising, super-resolution and image enhancement. Our \xnet achieves state-of-the-results on \textit{all} six datasets. Furthermore, we extensively evaluate our approach on practical challenges, such as generalization ability across datasets (Sec.~\ref{sec:exp})}

In Table~\ref{table:compute comparisons}, we compare \xnet with MIRNet~\cite{zamir2020mirnet} under the same training and inference settings. 
The results show that \xnet is more accurate (improving PSNR from 39.72 dB to 39.84 dB), while reducing the number of parameters and FLOPs by $\sim 81\%$, convolutions by $36\%$, and activations by $69\%$. Furthermore, the training and inference speed is increased by $2.2\times$ and $3.6\times$, respectively.

\begin{figure*}[t]
\begin{center}
\begin{tabular}[t]{c} \hspace{-2mm}
\includegraphics[width=\textwidth]{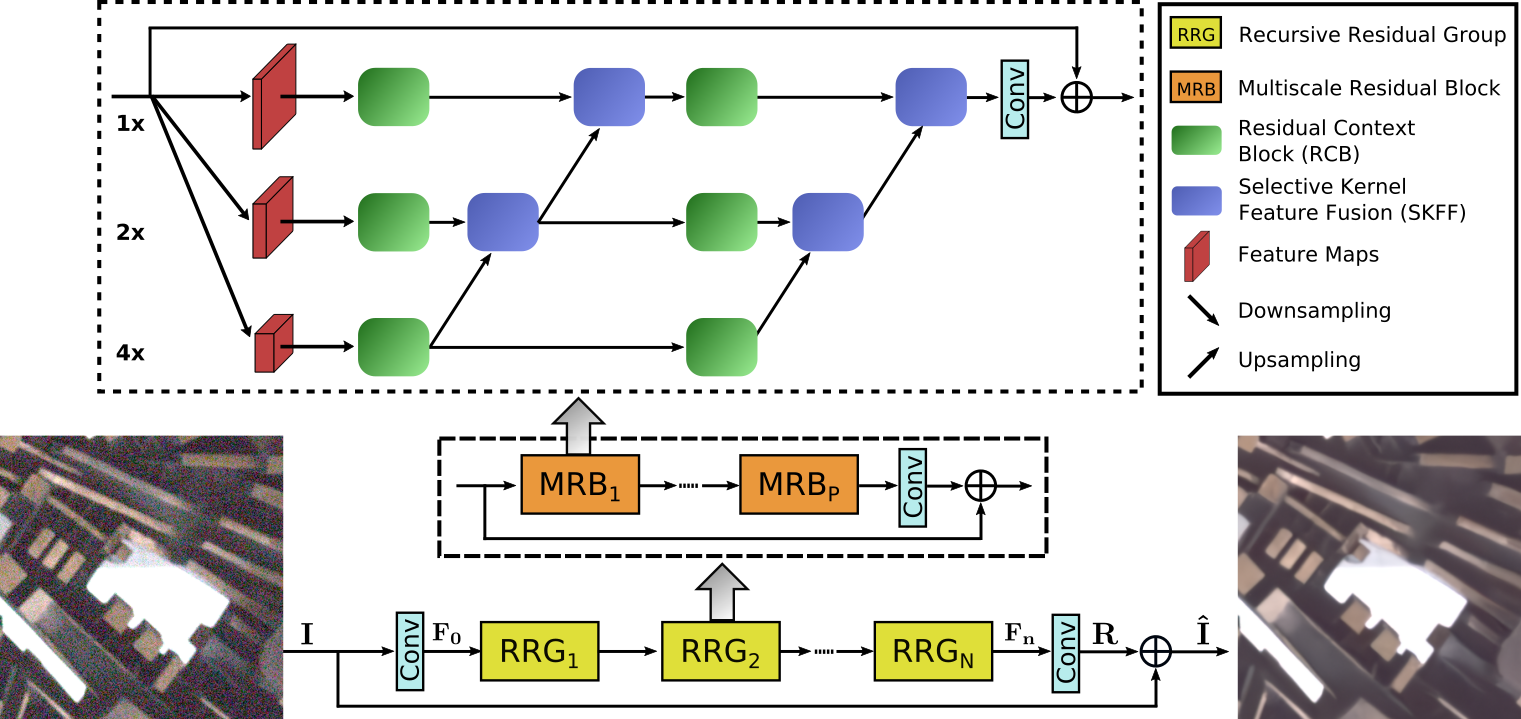}
\end{tabular}
\end{center}
\vspace*{-3mm}
\caption{Framework of the proposed \xnet that learns enriched feature representations for image restoration and enhancement. \xnet is based on a recursive residual design. In the core of \xnet is the multi-scale residual block (MRB) whose main branch is dedicated to maintaining spatially-precise high-resolution representations through the entire network and the complimentary set of parallel branches provide better contextualized features. 
}
\label{fig:framework}
\end{figure*}

\section{Related Work}
Rapidly growing image content necessitates the need to develop effective image restoration and enhancement algorithms. 
In this paper, we propose a new method capable of performing dual-pixel defocus deblurring, image denoising, super-resolution, and image enhancement.
{Unlike existing works for these problems, our approach processes features at the original resolution in order to preserve spatial details, while effectively fuses contextual information from multiple parallel branches.} 
Next, we briefly describe the representative methods for each of the studied problems.


\subsection{Dual-Pixel Defocus Deblurring} 
Images captured with wide camera aperture have shallow depth of field (DoF), where the scene regions that lie outside the DoF are out-of-focus. 
Given an image with defocus blur, the goal of defocus deblurring is to generate an all-in-focus image. Existing defocus deblurring approaches either directly deblur images~\cite{dandres2016,abdullah2020dpdd,abdullah2021rdpd,zamir2021restormer}, or first estimate the defocus dispartiy map and then use it to guide the deblurring procedure~\cite{karaali2017edge_EBDB,shi2015just_jnb,lee2019deep_dmenet}. 
Modern cameras are equipped with dual-pixel sensor that has two photodiodes at each pixel location, thereby generating two sub-aperture views. 
The phase difference between these views is useful in measuring the amount of defocus blur at each scene point. 
Recently, Abuolaim \textit{et al.}~\cite{abdullah2020dpdd} presented a dual-pixel deblurring dataset (DPDD) and a new method based on encoder-decoder design.
In this paper our focus is also on deblurring images directly using the dual-pixel data as in~\cite{abdullah2020dpdd,abdullah2021rdpd}.
{Previous defocus deblurring works~\cite{abdullah2020dpdd,abdullah2021rdpd} employ the encoder-decoder that repeatedly uses the downsampling operation, thus causing significant fine detail loss. Whereas the architectural design of our approach enables preservation of desired textural details in the restored image. }


\subsection{Image Denoising} 
Classic denoising methods are mainly based on modifying transform coefficients \cite{yaroslavsky1996local,simoncelli1996noise} or averaging neighborhood pixels \cite{tomasi1998bilateral,perona1990scale,rudin1992nonlinear}. 
Although the classical approaches perform well, the self-similarity \cite{efros1999texture} based algorithms, \textit{e.g.}, NLM \cite{NLM} and BM3D \cite{BM3D}, demonstrate promising denoising performance. 
Numerous patch-based schemes that exploit redundancy (self-similarity) in images are later developed \cite{dong2012nonlocal,WNNM,mairal2009non,hedjam2009markovian}. 
Recently, deep learning models \cite{RIDNet,Brooks2019,CBDNet,N3Net,DnCNN,FFDNetPlus,Zamir2020CycleISP,zamir2020mirnet,chang2020sadnet,yue2020danet,VDN,kim2020aindnet,fang2020multilevel_mlefgn,zamir2021multi} make significant advances in image denoising, yielding favorable results than those of the hand-crafted methods. 

%
\subsection{Image Super-Resolution} 
Prior to the deep-learning era, numerous super-resolution (SR) algorithms have been proposed based on the sampling theory~\cite{keys1981cubic,irani1991improving}, edge-guided interpolation \cite{allebach1996edge,zhang2006edge}, natural image priors \cite{kim2010single,xiong2010robust}, patch-exemplars \cite{chang2004super,freedman2011image} and sparse representations \cite{yang2010image,yang2008image}. 
Currently, deep-learning techniques are being actively explored as they provide dramatically improved results over conventional algorithms. 
The data-driven SR approaches differ according to their architecture designs~\cite{wang2019deep,anwar2019deep,ntire2019_superresolution}. 
Early methods~\cite{dong2014learning,dong2015image} take a low-resolution (LR) image as input and learn to directly generate its high-resolution (HR) version. 
In contrast to directly producing a latent HR image, recent SR networks \cite{VDSR,tai2017memnet,tai2017image,hui2018fast} employ the residual learning framework \cite{He2016} to learn the high-frequency image detail, which is later added to the input LR image to produce the final result.
Other networks designed to perform SR include recursive learning \cite{kim2016deeply,han2018image,ahn2018fast}, progressive reconstruction \cite{wang2015deep,Lai2017}, dense connections \cite{tong2017image,wang2018esrgan,zhang2020residual}, attention mechanisms \cite{RCAN,dai2019second,zhang2019residual}, multi-branch learning \cite{Lai2017,EDSR,dahl2017pixel,li2018multi}, and generative adversarial networks (GANs) \cite{wang2018esrgan,park2018srfeat,sajjadi2017enhancenet,SRResNet}. 
%
%

\subsection{Image Enhancement}
Oftentimes, cameras generate images that lack vivid details or contrast. 
A number of factors contribute to the low quality of images, including unsuitable lighting conditions and physical limitations of camera devices. 
For image enhancement, histogram equalization is the most commonly used approach. However, it frequently produces under- or over-enhanced images. %
Motivated by the Retinex theory \cite{land1977retinex}, several enhancement algorithms mimicking human vision have been proposed in the literature \cite{bertalmio2007,palma2008perceptually,jobson1997multiscale,rizzi2004retinex}.
Recently, CNNs have been successfully applied to general, as well as low-light, image enhancement problems \cite{ntire2019_enhancement}. 
Notable works employ Retinex-inspired networks \cite{Shen2017,wei2018deep,zhang2019kindling,chang2015retinex}, encoder-decoder networks \cite{chen2018encoder,Lore2017,ren2019low,mei2019higher,li2020luminance}, and GANs \cite{chen2018deep,ignatov2018wespe,deng2018aesthetic}.

\begin{figure*}[t]
\begin{center}
\scalebox{0.85}{
\begin{tabular}[t]{c} \hspace{-2mm}
\includegraphics[width=\textwidth]{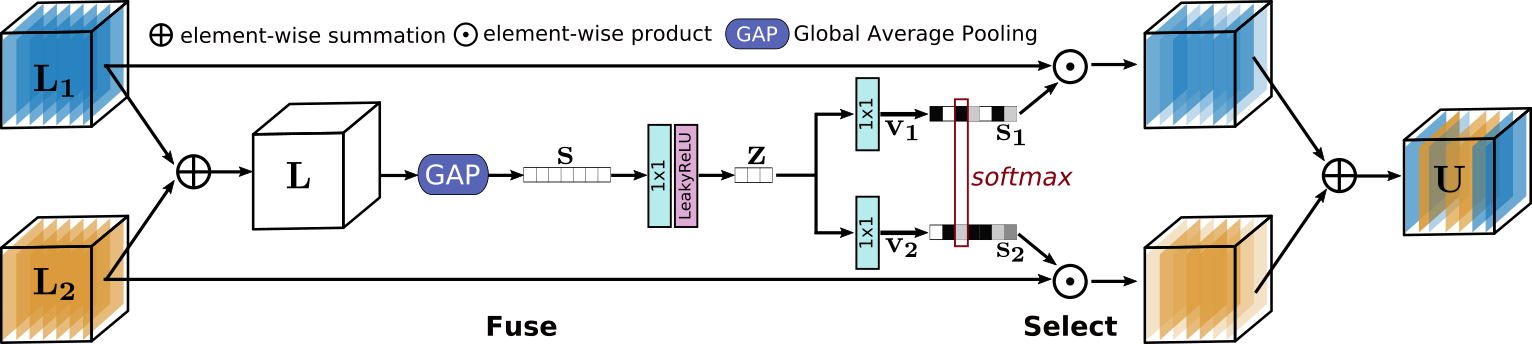}
\end{tabular}}
\end{center}
\vspace*{-3mm}
\caption{Schematic for selective kernel feature fusion (SKFF). It operates on features from different resolution streams, and performs aggregation based on self-attention.}
\label{fig:skff}
\end{figure*}


\section{Proposed Method}\label{sec:method}
A schematic of the proposed \xnet is shown in Fig.~\ref{fig:framework}. 
We first present an overview of the proposed \xnet for image restoration and enhancement.
We then provide details of the \emph{multi-scale residual block}, which is the fundamental building block of our method, containing several key elements: \textbf{(a)} parallel multi-resolution convolution streams for extracting (fine-to-coarse) semantically-richer and (coarse-to-fine) spatially-precise feature representations, \textbf{(b)} information exchange across multi-resolution streams, \textbf{(c)} attention-based aggregation of features arriving from different streams, and \textbf{(d)} residual contextual blocks to extract attention-based features.

\vspace{0.4em} \noindent \textbf{Overall Pipeline.} 
Given an image $\mathbf{I} \in \mathbb{R}^{H\times W \times 3}$, the proposed model first applies a convolutional layer to extract low-level features $\mathbf{F_0} \in \mathbb{R}^{H\times W \times C}$.
Next, the feature maps $\mathbf{F_0}$ pass through $N$ number of recursive residual groups (RRGs), yielding deep features $\mathbf{F_n} \in \mathbb{R}^{H\times W \times C}$. 
We note that each RRG contains several multi-scale residual blocks, which is described in Section~\ref{sec:msrb}. 
Next, we apply a convolution layer to deep features $\mathbf{F_n}$ and obtain a residual image $\mathbf{R} \in \mathbb{R}^{H\times W \times 3}$. 
Finally, the restored image is obtained as $\mathbf{\hat{I}} = \mathbf{I} + \mathbf{R}$.   
We optimize the proposed network using the Charbonnier loss \cite{charbonnier1994}: 
\begin{equation}
\label{Eq:loss}
\mathcal{L}(\mathbf{\hat{I}},\mathbf{I}^*) = \sqrt{ {\|\mathbf{\hat{I}}-\mathbf{I}^*\|}^2 + {\varepsilon}^2 },
\end{equation}
where $\mathbf{I}^*$ denotes the ground-truth image, and $\varepsilon$ is a constant which we empirically set to $10^{-3}$ for all the experiments.
%

\subsection{Multi-Scale Residual Block}
\label{sec:msrb}
To encode context, existing CNNs \cite{ronneberger2015u,newell2016stacked,noh2015learning,xiao2018simple,badrinarayanan2017segnet,peng2016recurrent} typically employ the following architecture design: \textbf{(a)} the receptive field of neurons is fixed in \textit{each} layer/stage, \textbf{(b)} the spatial size of feature maps is \textit{gradually} reduced to generate a semantically strong low-resolution representation, and \textbf{(c)} a high-resolution representation is \textit{gradually} recovered from the low-resolution representation.  
However, it is well-understood in vision science that in the primate visual cortex, the sizes of the local receptive fields of neurons in the same region are different \cite{hubel1962receptive,riesenhuber1999hierarchical,serre2007robust,hung2005fast}. 
Therefore, a similar mechanism of collecting multi-scale spatial information in the same layer is more effective when incorporated with in CNNs~\cite{huang2017multi,hrnet,fourure2017residual,Szegedy2015}.
Motivated by this, we propose the multi-scale residual block (MRB), as shown in Fig.~\ref{fig:framework}. 
It is capable of generating a spatially-precise output by maintaining high-resolution representations, while receiving rich contextual information from low-resolutions. 
The MRB consists of multiple (three in this paper) fully-convolutional streams connected in parallel that operate on varying resolution feature maps (ranging from low to high). 
It allows contextualized-information transfer from the low-resolution streams to consolidate the high-resolution features. Next, we describe the individual components of MRB.

\begin{figure*}[t]
\begin{center}
\scalebox{0.85}{
\begin{tabular}[t]{c} \hspace{-2mm}
\includegraphics[width=\textwidth]{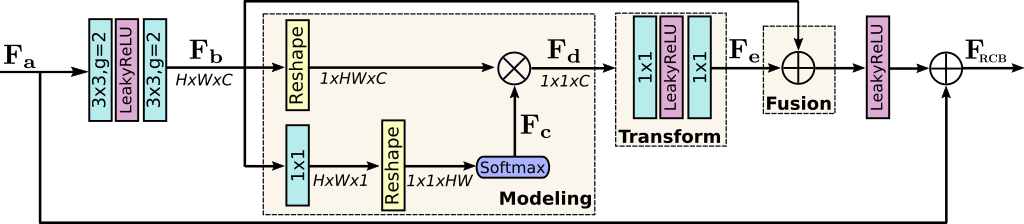}
\end{tabular}}
\end{center}
\vspace*{-3mm}
\caption{\small Architecture of residual contextual block (RCB). In the first two group convolution layers, $g$ represents the number of groups. $\otimes$ denotes matrix multiplication.}
\label{fig:rcb}
\end{figure*}

\subsubsection{Selective Kernel Feature Fusion}\label{sec:skff}
One fundamental property of neurons present in the visual cortex is their ability to change receptive fields according to the stimulus \cite{li2019selective}. 
This mechanism of adaptively adjusting receptive fields can be incorporated in CNNs by using multi-scale feature generation (in the same layer) followed by feature aggregation and selection. 
The most commonly used approaches for feature aggregation include simple concatenation or summation. 
However, these choices provide limited expressive power to the network, as reported in \cite{li2019selective}.
In MRB, we introduce a nonlinear procedure for fusing features coming from different resolution streams using a self-attention mechanism. 
Motivated by \cite{li2019selective}, we call it selective kernel feature fusion (SKFF). 

The SKFF module performs dynamic adjustment of receptive fields via two operations -- {\emph{Fuse} and \emph{Select}, as illustrated in Fig.~\ref{fig:skff}}.
The \emph{fuse} operator generates global feature descriptors by combining the information from multi-resolution streams. %
The \emph{select} operator uses these descriptors to recalibrate the feature maps (of different streams) followed by their aggregation. 
Next, we provide details of both operators.
\textbf{(1) Fuse:} SKFF receives inputs from two parallel convolution streams carrying different scales of information. 
We first combine these multi-scale features using an element-wise sum as: $\mathbf{L = L_1 + L_2}$. 
We then apply global average pooling (GAP) across the spatial dimension of $\mathbf{L} \in \mathbb{R}^{H\times W \times C}$ to compute channel-wise statistics $\mathbf{s} \in \mathbb{R}^{1\times 1 \times C}$. 
Next, we apply a channel-downscaling convolution layer to generate a compact feature representation $\mathbf{z} \in \mathbb{R}^{1\times 1 \times r}$, where $r=\frac{C}{8}$ for all our experiments.
Finally, the feature vector $\mathbf{z}$ passes through two parallel channel-upscaling convolution layers (one for each resolution stream) and provides us with two feature descriptors $\mathbf{v_1}$ and $\mathbf{v_2}$, each with dimensions $1\times1\times C$. 
\textbf{(2) Select:} This operator applies the softmax function to $\mathbf{v_1}$ and $\mathbf{v_2}$, yielding attention activations $\mathbf{s_1}$ and $\mathbf{s_2}$ that we use to adaptively recalibrate multi-scale feature maps $\mathbf{L_1}$ and $\mathbf{L_2}$, respectively. 
The overall process of feature recalibration and aggregation is defined as: $\mathbf{U = s_1 \cdot L_1 + s_2\cdot L_2}$. 
Note that the SKFF uses $\sim$$5$x fewer parameters than aggregation with concatenation but generates more favorable results (an ablation study is provided in the experiments section).


\subsubsection{Residual Contextual Block}\label{sec:rcb}
{While the SKFF block fuses information across multi-resolution branches, we also need a distillation mechanism to extract useful information from within a feature tensor.}
Motivated by the advances of recent low-level vision methods ~\cite{RCAN,RIDNet,dai2019second,zhang2019residual} which incorporate attention mechanisms~\cite{hu2018squeeze,wang2018non,cao2020global,khan2021transformers}, we propose the residual contextual block (RCB) to extract features in the convolutional streams. The schematic of RCB is shown in Fig.~\ref{fig:rcb}.
The RCB suppresses less useful features and only allows more informative ones to pass further. 
The overall process of RCB is summarized as:
\begin{align}
\mathbf{F_{RCB}} = \mathbf{F_{a}} + W(\text{CM}(\mathbf{F_b})),
\label{Eq:rcb}
\end{align}
where $\mathbf{F_b} \in \mathbb{R}^{H\times W \times C}$ represents feature maps that are obtained by applying two $3$x$3$ \textit{group} convolution layers to the input features $\mathbf{F_b} \in \mathbb{R}^{H\times W \times C}$ at the beginning of the RCB. These group convolutions are more resource efficient than standard convolutions and capable of learning unique representations in each filter group. $W$ denotes the last convolutional layer with filter size $1$x$1$. CM stands for contextual module that is realized in three parts. 
\textbf{(1) Context modeling:} From the original feature maps $\mathbf{F_b}$, we first generate new features $\mathbf{F_c} \in \mathbb{R}^{1\times 1 \times HW}$ by applying 1x1 convolution followed by the reshaping and softmax operations. 
Next we reshape $\mathbf{F_b}$ to $\mathbb{R}^{1\times HW \times C}$ and perform matrix multiplication with $\mathbf{F_c}$ to obtain the global feature descriptor $\mathbf{F_d} \in \mathbb{R}^{1\times 1 \times C}$. %
\textbf{(2) Feature transform:} To capture the inter-channel dependencies we pass the descriptor $\mathbf{F_d}$ through two 1x1 convolutions, resulting in new attention features $\mathbf{F_e} \in \mathbb{R}^{1\times 1 \times C}$.
\textbf{(3) Feature fusion:} We employ element-wise addition operation to aggregate contextual features $\mathbf{F_e}$ to each position of the original features $\mathbf{F_b}$.
%


\subsection{Progressive Training Regime}\label{sec:progressive}
When considering the image patch size for network training, there is a trade-off between the training speed and test-time accuracy~\cite{hoffer2019mix,tan2019efficientnet}. 
On large patches, CNNs capture fine image details to provide improved results, but they are slower to train. Whereas, training on small image patches is faster, but comes at the cost of accuracy drop. To strike the right balance between the training speed and accuracy, we propose a progressive learning method where the network is trained on smaller image patches in the early epochs and on gradually larger patches in the later training epochs. This approach can also be understood as a curriculum learning process where the network sequentially moves from learning a simpler task to a more complex one (where modeling of fine details is required). 
The progressive learning strategy on mixed-size image patches not only improves the training speed but also enhances the model performance at test time where the input images can be of different sizes (which is common in image restoration problems). 


\section{Experiments}\label{sec:exp}
In this section, we perform qualitative and quantitative assessments of the results produced by our \xnet and compare it with the state-of-the-art methods. 
Next, we describe the datasets, and then provide the implementation details. Finally, we report results for \textbf{(a)} dual-pixel defocus deblurring, \textbf{(b)} image denoising, \textbf{(c)} image super-resolution and \textbf{(d)} image enhancement, on six real image datasets. 

\begin{table*}[!t]
\begin{center}
\caption{Dual-pixel Defocus Deblurring comparisons on the DPDD Dataset~\cite{abdullah2020dpdd}. The test set of DPDD contains 37 indoor scenes and 39 outdoor scenes. Best and second best scores are \textbf{highlighted} and \underline{underlined}, respectively.
}
\label{table:dpdd}
\setlength{\tabcolsep}{4pt}
\scalebox{1}{
\begin{tabular}{l | c c c c | c c c c | c c c c }
\toprule[0.15em]
   & \multicolumn{4}{c|}{Indoor Scenes} & \multicolumn{4}{c|}{Outdoor Scenes} & \multicolumn{4}{c}{Combined} \\
\cline{2-13}
   Method & PSNR~$\textcolor{black}{\uparrow}$ & SSIM~$\textcolor{black}{\uparrow}$& MAE~$\textcolor{black}{\downarrow}$ & LPIPS~$\textcolor{black}{\downarrow}$  & PSNR~$\textcolor{black}{\uparrow}$ & SSIM~$\textcolor{black}{\uparrow}$& MAE~$\textcolor{black}{\downarrow}$ & LPIPS~$\textcolor{black}{\downarrow}$  & PSNR~$\textcolor{black}{\uparrow}$ & SSIM~$\textcolor{black}{\uparrow}$& MAE~$\textcolor{black}{\downarrow}$ & LPIPS~$\textcolor{black}{\downarrow}$   \\
\midrule[0.15em]
EBDB~\cite{karaali2017edge_EBDB} & 25.77 & 0.772 & 0.040 & 0.297 & 21.25 & 0.599 & 0.058 & 0.373 & 23.45 & 0.683 & 0.049 & 0.336 \\
DMENet~\cite{lee2019deep_dmenet}  & 25.50 & 0.788 & 0.038 & 0.298 & 21.43 & 0.644 & 0.063 & 0.397 & 23.41 & 0.714 & 0.051 & 0.349 \\
JNB~\cite{shi2015just_jnb} & 26.73 & 0.828 & 0.031 & 0.273 & 21.10 & 0.608 & 0.064 & 0.355 & 23.84 & 0.715 & 0.048 & 0.315 \\
DPDNet~\cite{abdullah2020dpdd} & 27.48 & \underline{0.849} & 0.029 & \underline{0.189} & \underline{22.90} & \underline{0.726} & \underline{0.052} & \underline{0.255} & 25.13 & \underline{0.786} & 0.041 & \underline{0.223} \\
RDPD~\cite{abdullah2021rdpd} & \underline{28.10} & 0.843 & \underline{0.027} & 0.210 & 22.82 & 0.704 & 0.053 & 0.298 & \underline{25.39} & 0.772 & \underline{0.040} & 0.255 \\
\midrule[0.1em]
\textbf{\xnet~(Ours)} & \textbf{28.96} & \textbf{0.881} & \textbf{0.024} & \textbf{0.154} & \textbf{23.59} &	\textbf{0.753} & \textbf{0.049} & \textbf{0.205} & \textbf{26.20} & \textbf{0.816} & \textbf{0.037} & \textbf{0.180} \\
\bottomrule[0.1em]
\end{tabular}}
\end{center}
\end{table*}


\subsection{Real Image Datasets}
\noindent \textbf{Dual-pixel defocus deblurring.} 
\noindent \textbf{DPDD~\cite{abdullah2020dpdd}} dataset contains 500 indoor/outdoor scenes captured with a DSLR camera. Each scene consists of two defocus blurred sub-aperture views captured with a wide camera aperture, and the corresponding all-in-focus ground truth image captured with a narrow aperture. The DDPD dataset is divided into 350 images for training, 74 images for validation and 76 images for testing.    

\vspace{0.4em} \noindent \textbf{Image denoising.} 
\noindent \textbf{(1) DND \cite{dnd}} consists of $50$ images captured with four consumer cameras. %
Since the images are of very high-resolution, the dataset providers extract $20$ crops of size $512\times512$ from each image, yielding $1000$ patches in total. All these patches are used for testing (as DND does not contain training or validation sets).  
The ground-truth noise-free images are not released publicly, therefore the image quality scores in terms of PSNR and SSIM can only be obtained through an online server \cite{dndwebsite}. 
\noindent \textbf{(2) SIDD \cite{sidd}} is collected with smartphone cameras. Due to the small sensor and high-resolution, the noise levels in smartphone images are much higher than those of DSLRs. 
SIDD contains $320$ image pairs for training and $1280$ for validation. 

\vspace{0.4em}\noindent \textbf{Super-resolution.} 
\noindent \textbf{RealSR~\cite{RealSR}} contains real-world LR-HR image pairs of the same scene captured by adjusting the focal-length of the cameras. 
RealSR has both indoor and outdoor images taken with two cameras. 
The number of training image pairs for scale factors $\times2$, $\times3$ and $\times4$ are $183$, $234$ and $178$, respectively. For each scale factor, $30$ test images are also provided in RealSR.


\vspace{0.4em}\noindent \textbf{Image enhancement.} 
\noindent \textbf{(1)~LoL~\cite{wei2018deep}} is created for low-light image enhancement problem. It provides 485 images for training and 15 for testing. Each image pair in LoL consists of a low-light input image and its corresponding well-exposed reference image.
\noindent \textbf{(2)~MIT-Adobe FiveK \cite{mit_fivek}} contains $5000$ images of various indoor and outdoor scenes captured with DSLR cameras in different lighting conditions. 
The tonal attributes of all images are manually adjusted by five different trained photographers (labelled as experts A to E). 
Similar to \cite{hu2018exposure,park2018distort,wang2019underexposed}, we also consider the enhanced images of expert C as the ground-truth. 
Moreover, the first 4500 images are used for training and the last 500 for testing.

\subsection{Implementation Details}
The proposed architecture is end-to-end trainable and requires no pre-training of sub-modules. 
We train four different networks for four different restoration tasks. For the dual-pixel defocus deblurring, we concatenate the left and right sub-aperture images and feed them as input to the network. The training parameters, common to all experiments, are the following. 
We use 4 RRGs, each of which further contains $2$ MRBs. The MRB has $3$ parallel streams with channel dimensions of $80, 120, 180$ at resolutions $1, \frac{1}{2}, \frac{1}{4}$, respectively. Each stream in MRB has $2$ RCBs with shared parameters.  
The models are trained with the Adam optimizer ($\beta_1 = 0.9$, and $\beta_2=0.999$) for $3\times10^5$ iterations. The initial learning rate is set to $2\times10^{-4}$. We employ the cosine annealing strategy~\cite{loshchilov2016sgdr} to steadily decrease the learning rate from initial value to $10^{-6}$ during training. 
For progressive training, we use the image patch sizes of 128, 144, 192, and 224. 
The batch size is set to $64$ and, for data augmentation, we perform horizontal and vertical flips.

\subsection{Dual-Pixel Defocus Deblurring}
We compare the performance of the proposed \xnet with the conventional defocus deblurring methods (EBDB~\cite{karaali2017edge_EBDB} and JNB~\cite{shi2015just_jnb}) as well as the learning-based approaches (DMENet~\cite{lee2019deep_dmenet}, DPDNet~\cite{abdullah2020dpdd}, and RDPD~\cite{abdullah2021rdpd}).
Table~\ref{table:dpdd} shows that our method achieves state-of-the-art results for both the indoor and outdoor scene categories. 
In particular, our \xnet achieves 0.86 dB PSNR improvement over the previous best method RDPD~\cite{abdullah2021rdpd} on indoor images and 0.77 dB on outdoor images. 
When both scene categories are combined, our method shows performance gains of 0.81 dB over RDPD~\cite{abdullah2020dpdd} and 1.07 dB over the second best method DPDNet~\cite{abdullah2020dpdd}. 

\begin{figure*}[!t]
\begin{center}
\scalebox{1}{
\begin{tabular}[b]{c@{ } c@{ }  c@{ } c@{ } c@{ }   }\hspace{-4mm}
\multirow{4}{*}{\includegraphics[width=.36\textwidth,valign=t]{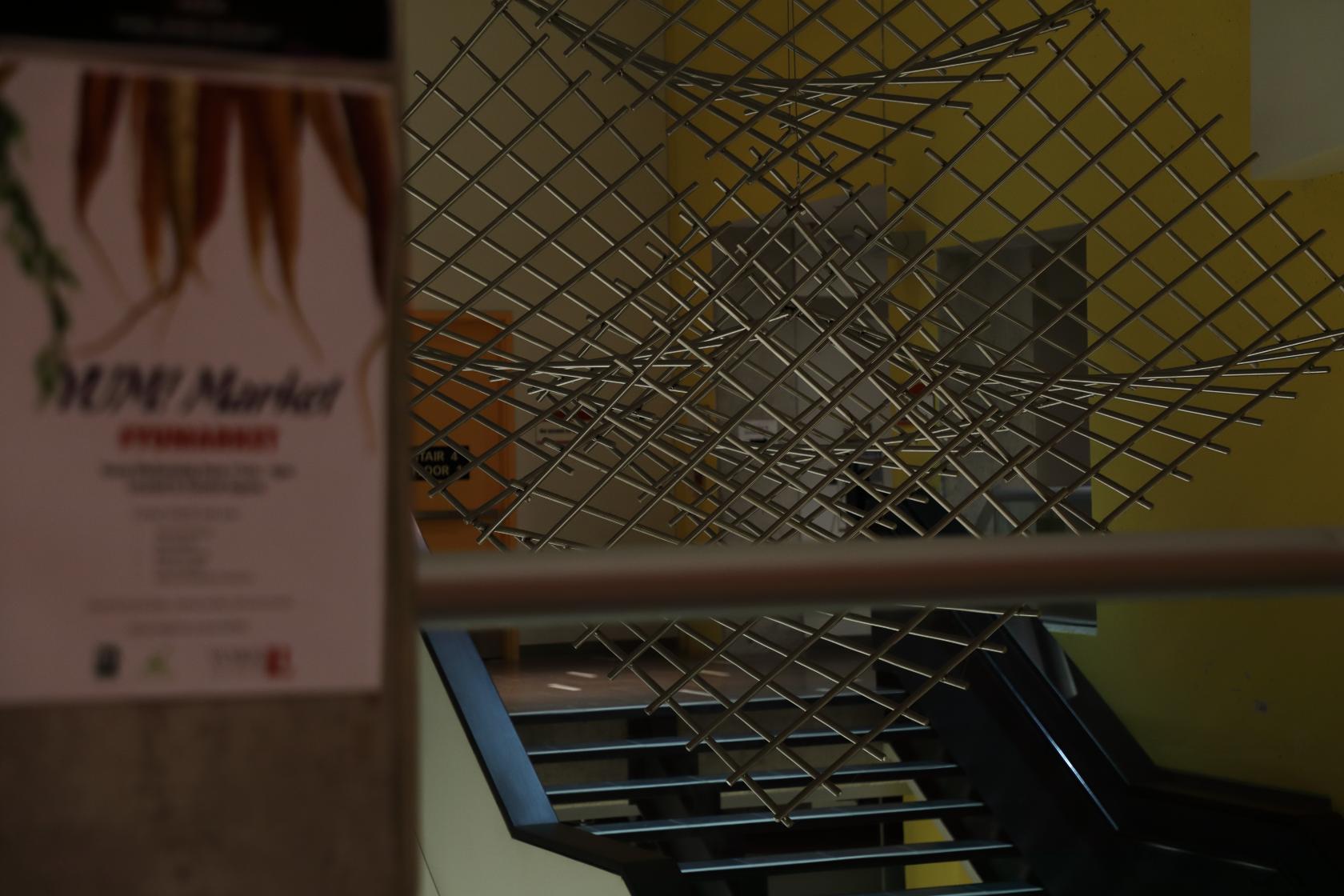}} &   
    \includegraphics[width=.145\textwidth,valign=t]{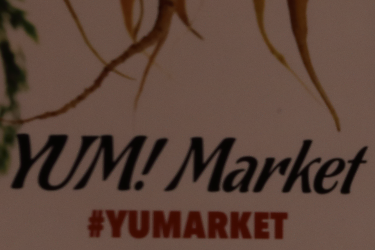} &
    \includegraphics[width=.145\textwidth,valign=t]{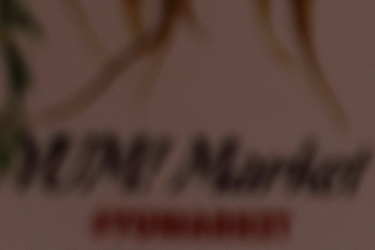} &
    \includegraphics[width=.145\textwidth,valign=t]{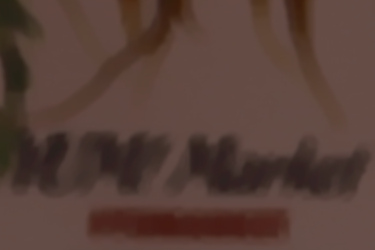} &
    \includegraphics[width=.145\textwidth,valign=t]{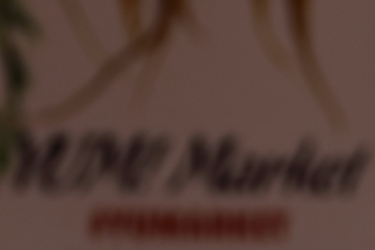}
\\
    &  \small~PSNR &\small~27.19 dB  & \small~26.43 dB & \small~27.44 dB   \\
    & \small~Reference & \small~Blurry  & \small EBDB~\cite{karaali2017edge_EBDB} & \small~DMENet~\cite{lee2019deep_dmenet}   \\

    &
    \includegraphics[width=.145\textwidth,valign=t]{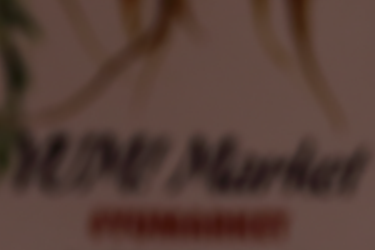} &
    \includegraphics[width=.145\textwidth,valign=t]{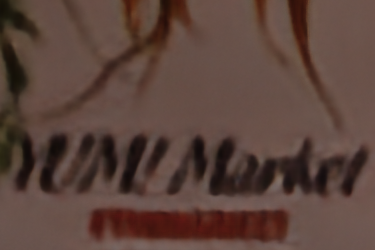} &
    \includegraphics[width=.145\textwidth,valign=t]{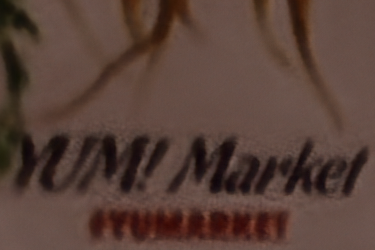} &
    \includegraphics[width=.145\textwidth,valign=t]{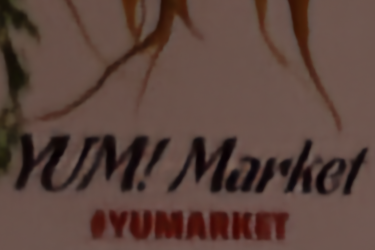}\\
     \small~27.19 dB& \small~26.82 dB & \small~28.67 dB 
     & \small~29.01 dB & \small~\textbf{29.82 dB}\\
           \small~Blurry Image  & \small JNB~\cite{shi2015just_jnb} & \small~DPDNet~\cite{abdullah2020dpdd} &  \small~RDPD~\cite{abdullah2021rdpd} &  \small~\textbf{\xnet}
\\
\end{tabular}}
\end{center}
\vspace*{-3mm}
\caption{Visual comparisons for dual-pixel defocus deblurring on the DPDD dataset~\cite{abdullah2020dpdd}. Compared to the other approaches, our \xnet more effectively removes blur while preserving the fine image details. 
}
\label{fig:dpdd results}
\end{figure*}

In Fig.~\ref{fig:dpdd results}, we provide defocus-deblurred results produced by different methods for both indoor and outdoor scenes. 
It is noticeable that our method effectively removes the spatially varying defocus blur and produces images that are more sharper and visually faithful to the ground-truth than those of the compared approaches.

\begin{table}[!t]
\begin{center}
\caption{Denoising comparisons on SIDD~\cite{sidd} and DND~\cite{dnd} datasets. \textcolor{red}{$\ast$} indicates the methods that use additional training data. Whereas our \xnet is only trained on the SIDD images and directly tested on DND.}
\label{table:denoising}
\setlength{\tabcolsep}{6.5pt}
\scalebox{0.99}{
\begin{tabular}{l | c c | c c }
\toprule[0.15em]
 & \multicolumn{2}{c|}{SIDD~\cite{sidd}} & \multicolumn{2}{c}{DND~\cite{dnd}} \\
 \cline{2-5}
 Method & PSNR~$\textcolor{black}{\uparrow}$ & SSIM~$\textcolor{black}{\uparrow}$ & PSNR~$\textcolor{black}{\uparrow}$ & SSIM~$\textcolor{black}{\uparrow}$\\
\midrule[0.15em]
DnCNN~\cite{DnCNN}                             & 23.66  & 0.583  & 32.43  & 0.790 \\
MLP~\cite{MLP}                                 & 24.71  & 0.641  & 34.23  & 0.833 \\
BM3D~\cite{BM3D}                               & 25.65  & 0.685  & 34.51  & 0.851 \\
CBDNet\textcolor{red}{*}~\cite{CBDNet}         & 30.78  & 0.801  & 38.06  & 0.942 \\
DAGL~\cite{mou2021dynamicDAGL}                 & 38.94  & 0.953  & \underline{39.77}  & 0.956 \\
RIDNet\textcolor{red}{*}~\cite{RIDNet}         & 38.71  & 0.951  & 39.26 & 0.953 \\
AINDNet\textcolor{red}{*}~\cite{kim2020aindnet} & 38.95 & 0.952  & 39.37 & 0.951 \\
VDN~\cite{VDN}                                  & 39.28 & 0.956 & 39.38 & 0.952 \\
DeamNet\textcolor{red}{*}~\cite{ren2021adaptivedeamnet}& 39.47 & 0.957 & 39.63 & 0.953 \\
SADNet\textcolor{red}{*}~\cite{chang2020sadnet} & 39.46 & 0.957 & 39.59 & 0.952 \\
DANet+\textcolor{red}{*}~\cite{yue2020danet}         & 39.47 & 0.957 & 39.58 & 0.955 \\
CycleISP\textcolor{red}{*}~\cite{Zamir2020CycleISP}  & \underline{39.52} & \underline{0.957} & 39.56 & \textbf{0.956} \\
\midrule[0.1em]
 \textbf{\xnet~(Ours)} & \textbf{39.84} & \textbf{0.959} & \textbf{39.86} 	& \underline{0.955} \\
\bottomrule[0.1em]
\end{tabular}}
\end{center}
\end{table}

\begin{figure*}[!t]
\begin{center}
\scalebox{1}{
\begin{tabular}[b]{c@{ } c@{ }  c@{ } c@{ } c@{ }   }\hspace{-4mm}
     \multirow{4}{*}{\includegraphics[width=.36\textwidth,valign=t]{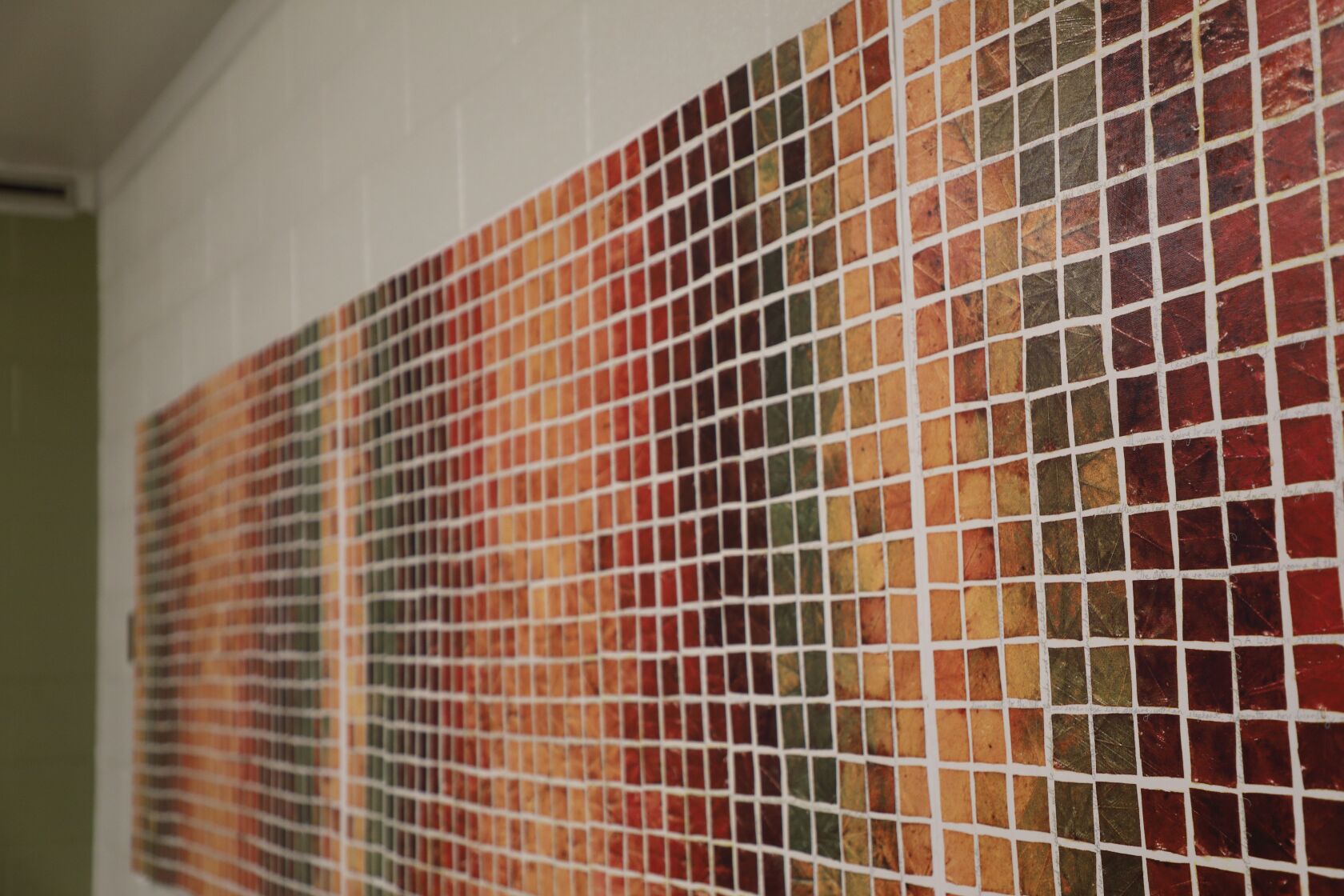}} &   
    \includegraphics[width=.145\textwidth,valign=t]{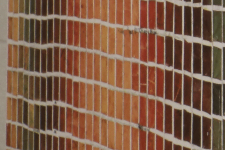} &
    \includegraphics[width=.145\textwidth,valign=t]{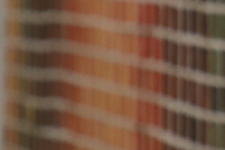} &
    \includegraphics[width=.145\textwidth,valign=t]{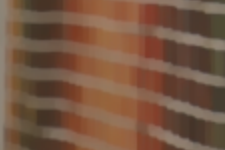}&
    \includegraphics[width=.145\textwidth,valign=t]{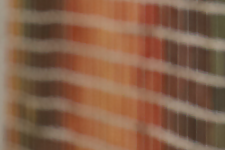}
\\
    &  \small~PSNR &\small~20.76 dB  & \small~21.02 dB & \small~21.37 dB   \\
    & \small~Reference & \small~Blurry  & \small EBDB~\cite{karaali2017edge_EBDB} & \small~DMENet~\cite{lee2019deep_dmenet}   \\

    &
    \includegraphics[width=.145\textwidth,valign=t]{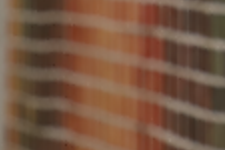} &
    \includegraphics[width=.145\textwidth,valign=t]{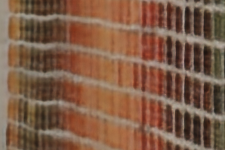} &
    \includegraphics[width=.145\textwidth,valign=t]{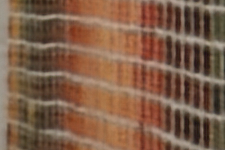} &
    \includegraphics[width=.145\textwidth,valign=t]{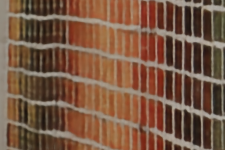}\\
     \small~20.76 dB & \small~20.96 dB  & \small~24.87 dB
     & \small~24.14 dB & \small~\textbf{26.41 dB}\\
           \small~Blurry Image  & \small JNB~\cite{shi2015just_jnb} & \small~DPDNet~\cite{abdullah2020dpdd} &  \small~RDPD~\cite{abdullah2021rdpd} &  \small~\textbf{\xnet}
\\
\end{tabular}}
\end{center}
\vspace{-6mm}
\end{figure*}

\begin{figure*}[!t]
\begin{center}
\scalebox{1}{
\begin{tabular}[b]{c@{ } c@{ }  c@{ } c@{ } c@{ }   }\hspace{-4mm}
\multirow{4}{*}{\includegraphics[width=.36\textwidth,valign=t]{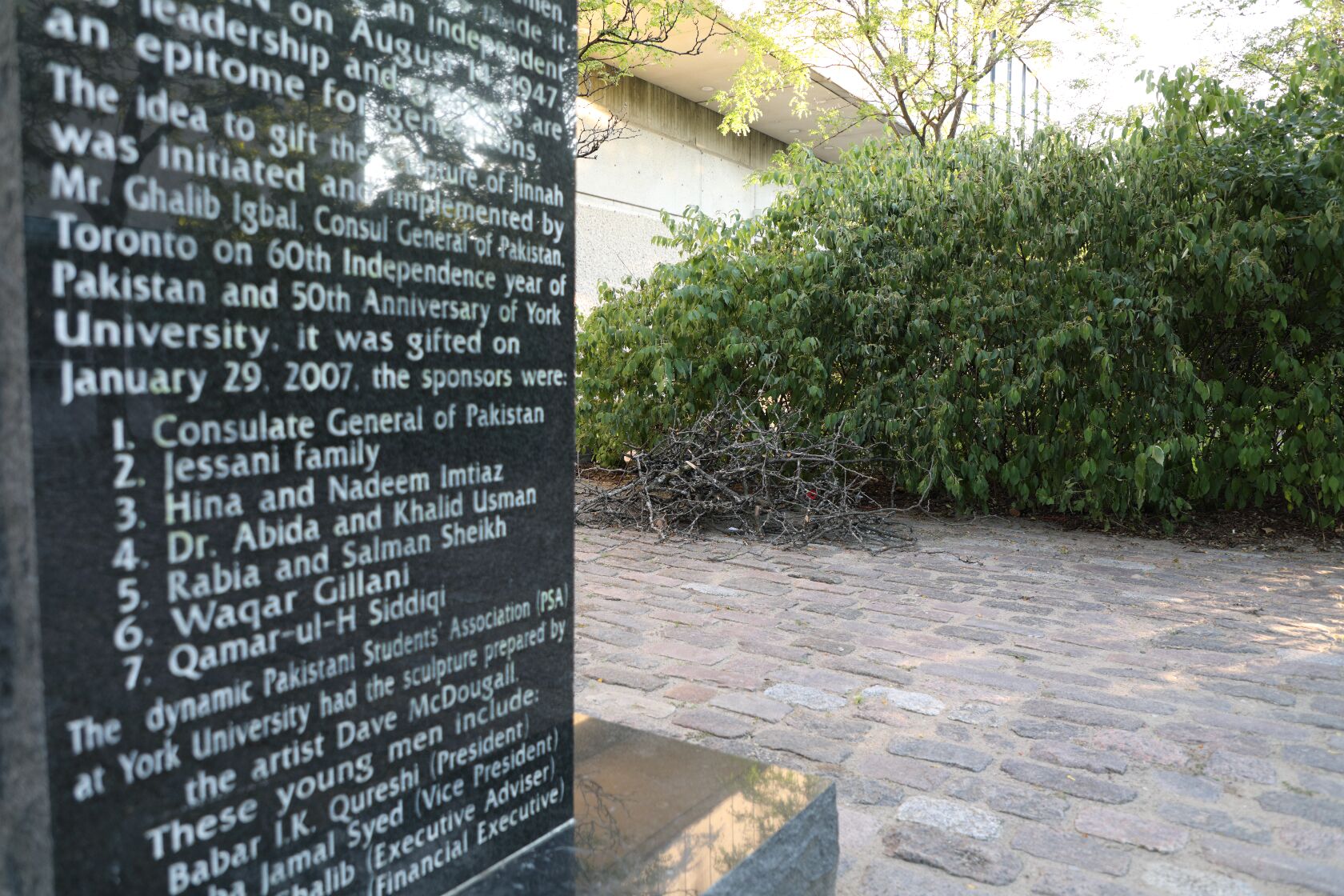}} &   
    \includegraphics[width=.145\textwidth,valign=t]{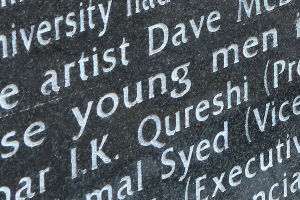} &
    \includegraphics[width=.145\textwidth,valign=t]{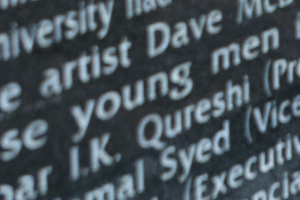} &
    \includegraphics[width=.145\textwidth,valign=t]{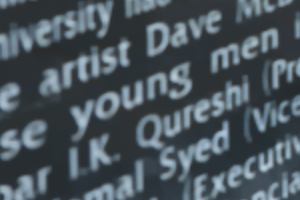} &
    \includegraphics[width=.145\textwidth,valign=t]{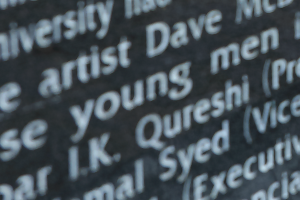}
\\
    &  \small~PSNR &\small~18.84 dB  & \small~18.64 dB & \small~18.75 dB   \\
    & \small~Reference & \small~Blurry  & \small EBDB~\cite{karaali2017edge_EBDB} & \small~DMENet~\cite{lee2019deep_dmenet}   \\

    &
    \includegraphics[width=.145\textwidth,valign=t]{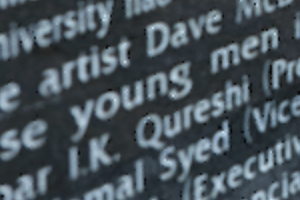} &
    \includegraphics[width=.145\textwidth,valign=t]{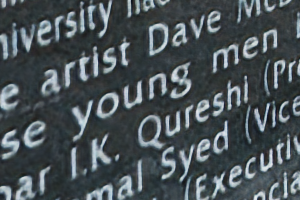} &
    \includegraphics[width=.145\textwidth,valign=t]{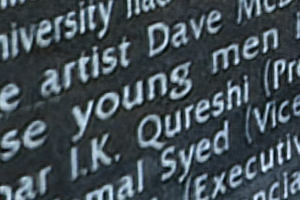} &
    \includegraphics[width=.145\textwidth,valign=t]{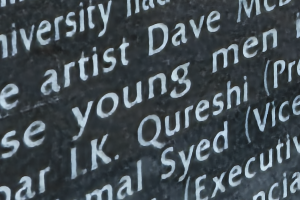}\\
     \small~18.84 dB& \small~18.65 dB & \small~20.01 dB 
     & \small~19.41 dB & \small~\textbf{20.69 dB}\\
           \small~Blurry Image  & \small JNB~\cite{shi2015just_jnb} & \small~DPDNet~\cite{abdullah2020dpdd} &  \small~RDPD~\cite{abdullah2021rdpd} &  \small~\textbf{\xnet}
\\
\end{tabular}}
\end{center}
\vspace{-6mm}
\end{figure*}

\subsection{Image Denoising}
In this section, we demonstrate the effectiveness of the proposed \xnet for image denoising.
We train our network only on the training set of the SIDD~\cite{sidd} and directly evaluate it on the test images of both SIDD and DND~\cite{dnd} datasets.
Quantitative comparisons in terms of PSNR and SSIM metrics are summarized in Table~\ref{table:denoising}.
Our \xnet performs favourably against the data-driven, as well as conventional, denoising algorithms. 
Specifically, when compared to the recent best methods, our algorithm demonstrates a performance gain of $0.32$~dB over CycleISP~\cite{Zamir2020CycleISP} on SIDD and $0.11$~dB over DAGL~\cite{mou2021dynamicDAGL} on DND.  
Furthermore, it is worth noting that CycleISP~\cite{Zamir2020CycleISP} uses additional training data, yet our method yields considerably better results.

\begin{figure*}[!t]
\begin{center}
\begin{tabular}[t]{c@{ }c@{ }c@{ }c@{ }c@{ }c@{ }c@{ }c}\hspace{-1mm}
\includegraphics[width=.12\textwidth]{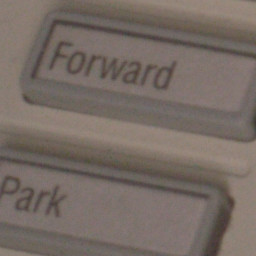}&   \hspace{-1.4mm}
\includegraphics[width=.12\textwidth]{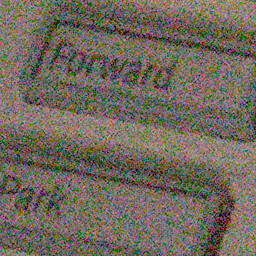}&    \hspace{-1.4mm}
\includegraphics[width=.12\textwidth]{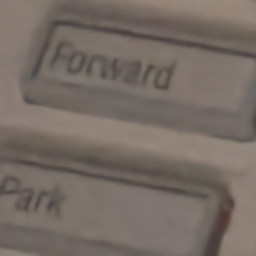}&   \hspace{-1.4mm}
\includegraphics[width=.12\textwidth]{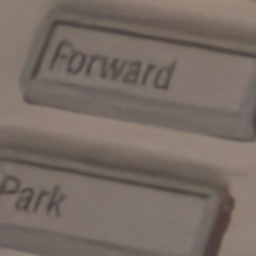}&   \hspace{-1.4mm}
\includegraphics[width=.12\textwidth]{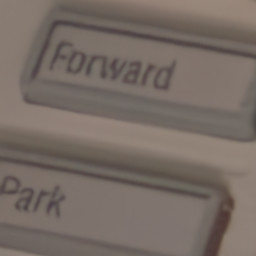}&   \hspace{-1.4mm}
\includegraphics[width=.12\textwidth]{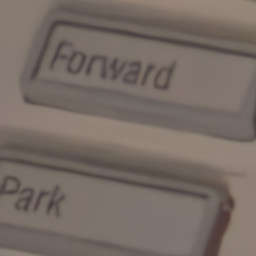}&   \hspace{-1.4mm}
\includegraphics[width=.12\textwidth]{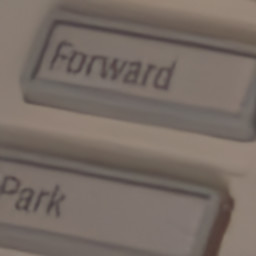}&   \hspace{-1.4mm}
\includegraphics[width=.12\textwidth]{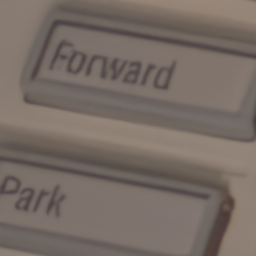}\\
\hspace{-1mm}\small~PSNR  &  \small~18.25 dB & \small~35.57 dB & \small~36.24 dB & \small~36.70 dB & \small~36.71 dB & \small~36.74 dB & \small~\textbf{37.07 dB}  \hspace{-4mm}\\
\includegraphics[width=.12\textwidth]{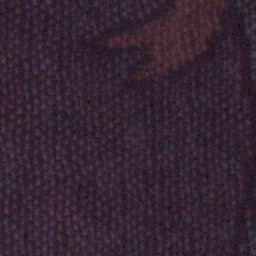}&   \hspace{-1.4mm}
\includegraphics[width=.12\textwidth]{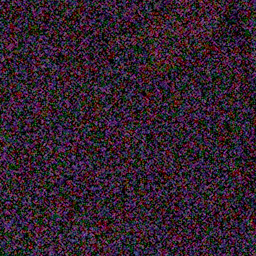}&    \hspace{-1.4mm}
\includegraphics[width=.12\textwidth]{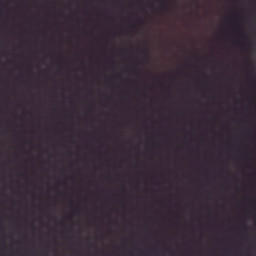}&   \hspace{-1.4mm}
\includegraphics[width=.12\textwidth]{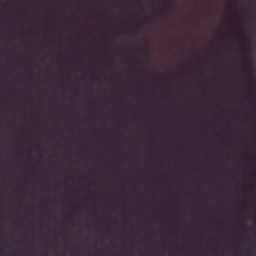}&   \hspace{-1.4mm}
\includegraphics[width=.12\textwidth]{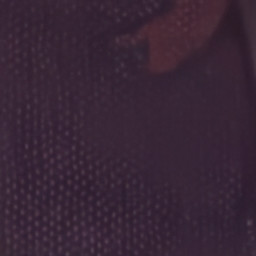}&   \hspace{-1.4mm}
\includegraphics[width=.12\textwidth]{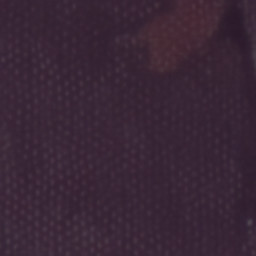}&   \hspace{-1.4mm}
\includegraphics[width=.12\textwidth]{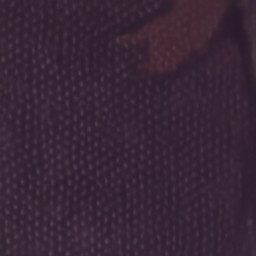}&   \hspace{-1.4mm}
\includegraphics[width=.12\textwidth]{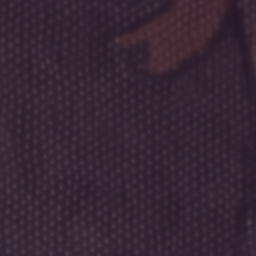}\\
\hspace{-1mm}\small~PSNR  &  \small~18.16 dB & \small~29.83 dB & \small~29.99 dB &  \small~30.48 dB & \small~30.22 dB & \small~30.76 dB & \small~\textbf{31.29 dB}  \\
\hspace{-1mm} \small Reference  & \small Noisy & \small RIDNet~\cite{RIDNet} & \small AINDNet~\cite{kim2020aindnet}  &\small  SADNet~\cite{chang2020sadnet} & \small CycleISP~\cite{Zamir2020CycleISP} & \small DANet~\cite{yue2020danet} &\textbf{\xnet}  \hspace{-2mm}
\end{tabular}
\vspace{3mm}
\scalebox{0.98}{
\begin{tabular}[b]{c@{ } c@{ }  c@{ } c@{ } c@{ } c@{ }	}\hspace{-4mm}
    \multirow{4}{*}{\includegraphics[width=.312\textwidth,valign=t]{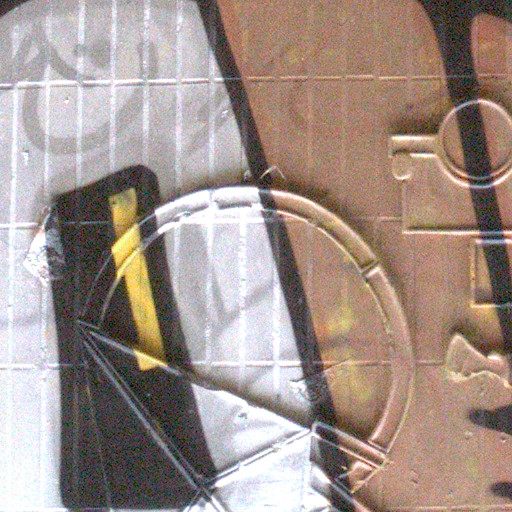}} &   
    \includegraphics[trim={10.8cm 6.5cm  3cm  7.2cm },clip,width=.13\textwidth,valign=t]{Images/Denoising/DND/RGB/noisy_26_90.jpg}&
  	\includegraphics[trim={10.8cm 6.5cm  3cm  7.2cm },clip,width=.13\textwidth,valign=t]{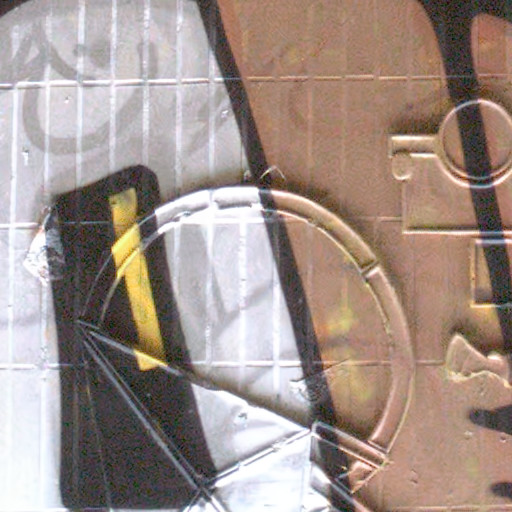}&   
    \includegraphics[trim={10.8cm 6.5cm  3cm  7.2cm },clip,width=.13\textwidth,valign=t]{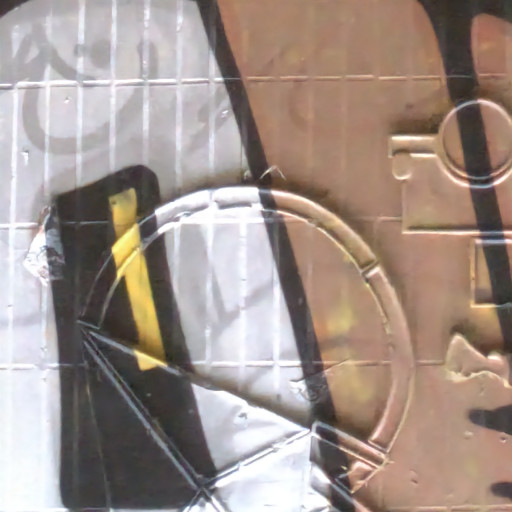}&
      	\includegraphics[trim={10.8cm 6.5cm  3cm  7.2cm },clip,width=.13\textwidth,valign=t]{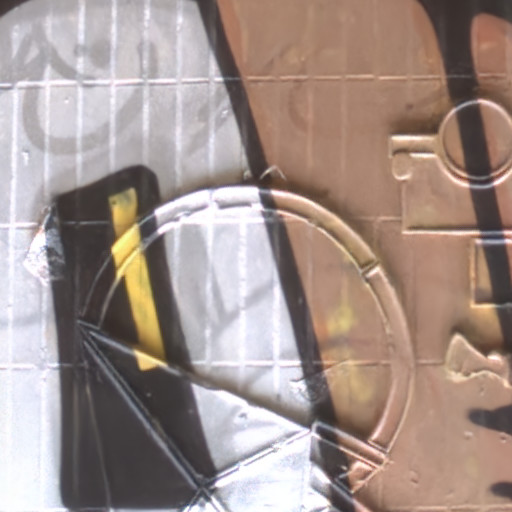}&
      \includegraphics[trim={10.8cm 6.5cm  3cm  7.2cm },clip,width=.13\textwidth,valign=t]{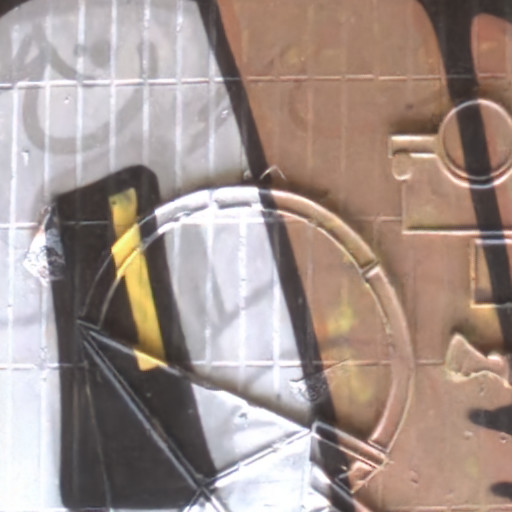}
  
\\
    &  \small~26.90 dB &\small~30.91 dB  & \small~33.62 dB & \small~33.89 dB & \small~34.09 dB   \\
    & \small~Noisy & \small~BM3D~\cite{BM3D}  & \small~CBDNet~\cite{CBDNet}  & \small~VDN~\cite{VDN} & \small~RIDNet~\cite{RIDNet} \\

    &
    \includegraphics[trim={10.8cm 6.5cm  3cm  7.2cm },clip,width=.13\textwidth,valign=t]{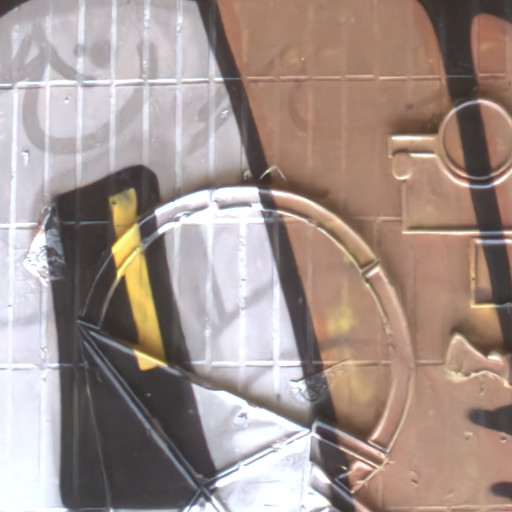}&
    \includegraphics[trim={10.8cm 6.5cm  3cm  7.2cm },clip,width=.13\textwidth,valign=t]{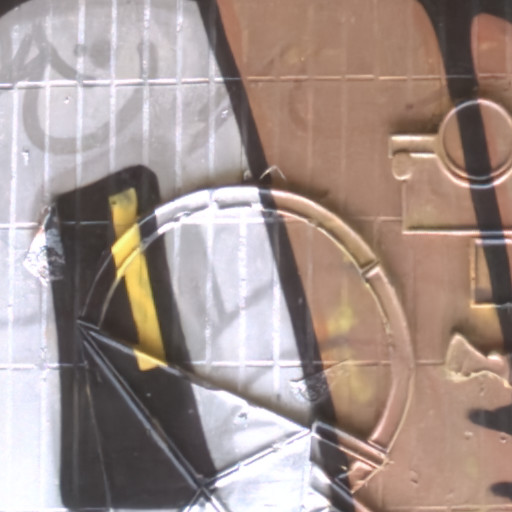}&
    \includegraphics[trim={10.8cm 6.5cm  3cm  7.2cm },clip,width=.13\textwidth,valign=t]{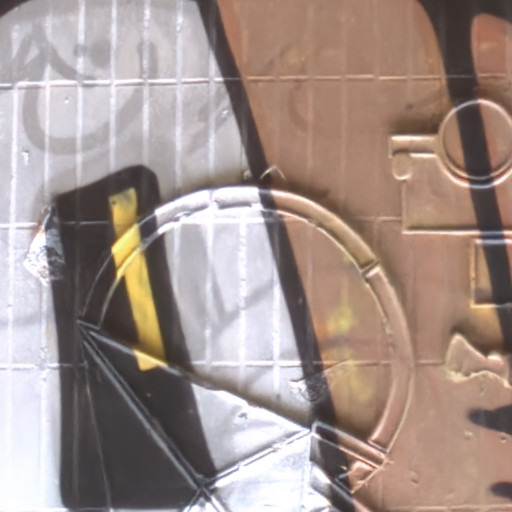}&  
     \includegraphics[trim={10.8cm 6.5cm  3cm  7.2cm },clip,width=.13\textwidth,valign=t]{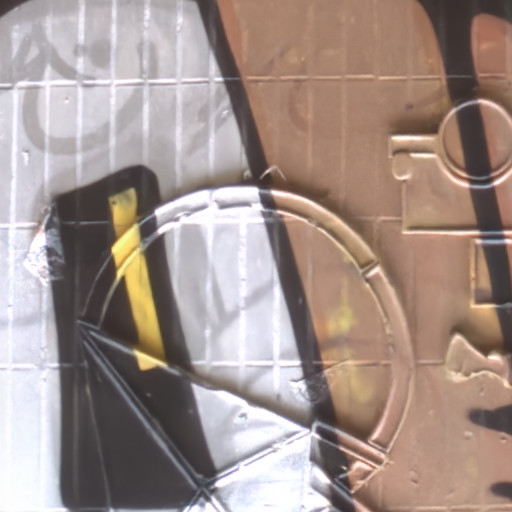}&
     \includegraphics[trim={10.8cm 6.5cm  3cm  7.2cm },clip,width=.13\textwidth,valign=t]{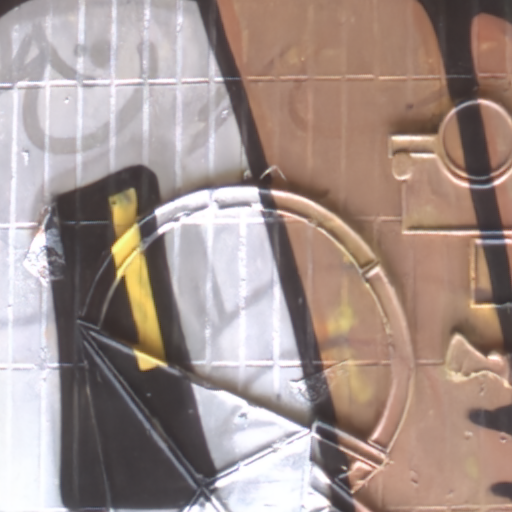}\\

     \small~26.90 dB& \small~34.32 dB& \small~34.36 dB & \small~34.36 dB 
     & \small~34.52 dB & \small~\textbf{34.64 dB}\\
           \small~Noisy Image  & \small~CycleISP~\cite{Zamir2020CycleISP}& \small~AINDNet~\cite{kim2020aindnet} & \small~DANet~\cite{yue2020danet} & \small~SADNet~\cite{chang2020sadnet}   & \small~\textbf{\xnet}
\\
\end{tabular}}
\end{center}
\vspace*{-5mm}
\caption{Image denoising comparisons. First two examples are from SIDD~\cite{sidd} and the last is from DND~\cite{dnd}. The proposed \xnet better preserves fine texture and structural patterns in the denoised images.
}
\label{fig:denoising}
\end{figure*}

Fig.~\ref{fig:denoising} shows a visual comparisons of our results with those of other competing algorithms. 
The \xnet is effective in removing real noise and produces perceptually-pleasing and sharp images.
Moreover, it is can maintain the spatial smoothness of the homogeneous regions without introducing artifacts.
In contrast, most of the other methods either yield over-smooth images and thus sacrifice structural content and fine textural details, or produce images with chroma artifacts and blotchy texture.

\vspace{0.4em}\noindent \textbf{Generalization capability.} The DND and SIDD datasets are acquired with different sets of cameras having different noise characteristics. Since the DND benchmark does not provide training data, setting a new state-of-the-art on DND with our SIDD trained network indicates the good generalization capability of our approach.


\subsection{Super-Resolution}

We compare our \xnet against the state-of-the-art SR algorithms (VDSR~\cite{VDSR}, SRResNet~\cite{SRResNet}, RCAN~\cite{RCAN}, LP-KPN~\cite{RealSR}) on the testing images of the RealSR~\cite{RealSR} for upscaling factors of $\times2$, $\times3$ and $\times4$. 
Note that all the benchmarked algorithms are trained on the RealSR~\cite{RealSR} dataset for a fair comparison. 
In the experiments, we also include bicubic interpolation~\cite{keys1981cubic}, which is the most commonly used method for generating super-resolved images. 
Here, we compute the PSNR and SSIM scores using the Y channel (in YCbCr color space), as it is a common practice in the SR literature~\cite{RCAN,RealSR,wang2019deep,anwar2019deep}. 
The results in Table~\ref{table:realSR} show that the bicubic interpolation provides the least accurate results, thereby indicating its low suitability for dealing with real images. 
Moreover, the same table shows that the recent method LP-KPN~\cite{RealSR} achieves marginal improvement of only $\sim0.04$~dB over the previous best method RCAN~\cite{RCAN}.
In contrast, our method significantly advances state-of-the-art and consistently achieves  better image quality scores than other approaches for all three scaling factors.
Particularly, compared to LP-KPN~\cite{RealSR}, our method leads to performance gains of $0.48$~dB, $0.73$~dB, and $0.24$~dB for scaling factors $\times2$, $\times3$ and $\times4$, respectively. The trend is similar for the SSIM metric as well.   

\begin{table}[t]
\begin{center}
\caption{Super-resolution evaluation on the RealSR dataset~\cite{RealSR} . Compared to the state-of-the-art, our method consistently yields significantly better image quality scores for all three scaling factors.}
\label{table:realSR}
\setlength{\tabcolsep}{4.5pt}
\scalebox{1}{
\begin{tabular}{l | c c | c c | c c}
\toprule[0.15em]
\multicolumn{1}{r|}{Scale} & \multicolumn{2}{c|}{x2} & \multicolumn{2}{c|}{x3} & \multicolumn{2}{c}{x4} \\
 \cline{2-7}
 Method & PSNR & SSIM  & PSNR & SSIM & PSNR & SSIM\\
\midrule[0.15em]
Bicubic &   32.61  &  0.907 &   29.34 &  0.841 &   27.99  &  0.806 \\
VDSR~\cite{VDSR} &   33.64  &  0.917 &   30.14  & 0.856 &  28.63  &  0.821 \\
SRResNet~\cite{SRResNet} &   33.69  &  0.919 &   30.18  &  0.859 &   28.67  &  0.824 \\
RCAN~\cite{RCAN} &  33.87  & 0.922 &  30.40  & 0.862  &  28.88  & 0.826 \\
LP-KPN~\cite{RealSR} &  \underline{33.90}  &  \underline{0.927}  &  \underline{30.42}  &  \underline{0.868}  & \underline{28.92}  &  \underline{0.834} \\ 
\midrule[0.1em]
\textbf{\xnet (Ours) }&   \textbf{34.38}  & \textbf{0.934} &   \textbf{31.15}  &  \textbf{0.883} &    \textbf{29.16}  &  \textbf{0.845}\\
\bottomrule[0.1em]
\end{tabular}}
\end{center}
\vspace{-1em}
\end{table}

\begin{figure*}[!h]
\begin{center}
\scalebox{1}{
\begin{tabular}[b]{c@{ } c@{ }  c@{ } c@{ } c@{ }	}\hspace{-1.5mm}
    \multirow{3}{*}{\includegraphics[width=.333\textwidth,valign=t]{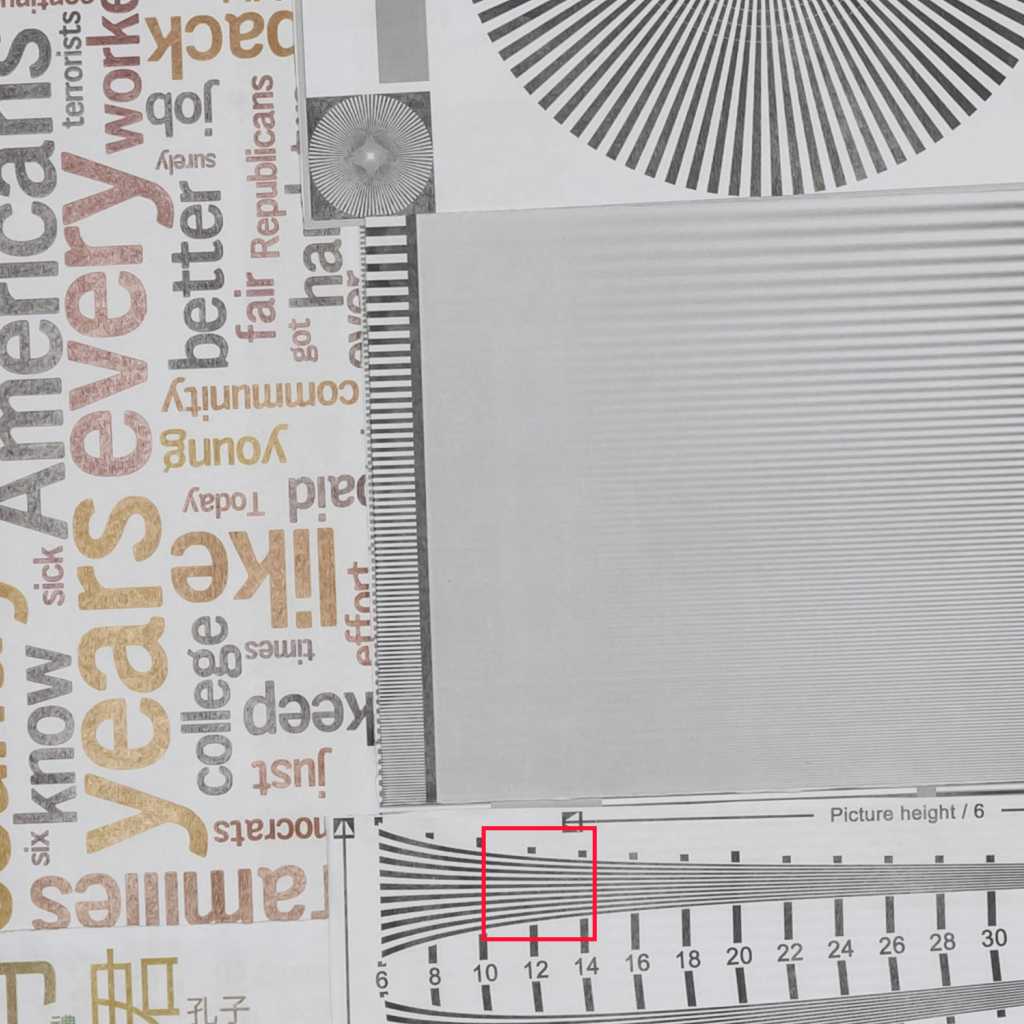}} 
    &  \includegraphics[width=.154\textwidth,valign=t]{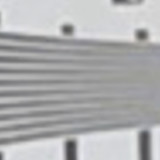}
    & \includegraphics[width=.154\textwidth,valign=t]{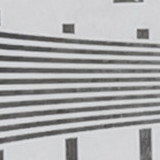}
    & \includegraphics[width=.154\textwidth,valign=t]{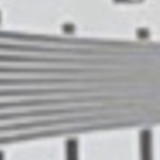}
    &\includegraphics[width=.154\textwidth,valign=t]{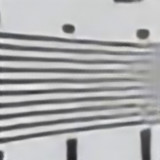}
    \vspace{0.5mm}
    \\
    \vspace{0.5mm}
    & \small{LR} & \small{HR}  & \small{Bicubic} & \small{SRResNet~\cite{SRResNet}} 
    \\
    &\includegraphics[width=.154\textwidth,valign=t]{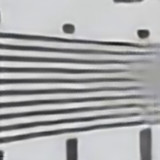}  
    &\includegraphics[width=.154\textwidth,valign=t]{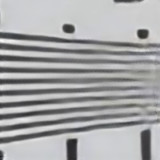}
    &\includegraphics[width=.154\textwidth,valign=t]{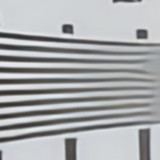}
    &\includegraphics[width=.154\textwidth,valign=t]{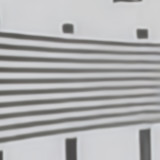}
    \vspace{0.5mm}
    \\
    \small{Image}  & \small{VDSR~\cite{VDSR}} & \small{RCAN~\cite{RCAN}} & \small{LP-KPN~\cite{RealSR}} & \small{\xnet (Ours)} \\
\end{tabular}}
\end{center}
\vspace{-3mm}
\caption{Comparisons for $\times4$ super-resolution on the RealSR~\cite{RealSR} dataset. The image produced by our \xnet is more faithful to the ground-truth than other competing methods (see lines near the right edge of the crops).}
\label{fig:sr example}
\end{figure*}

\begin{figure*}[t]
\begin{center}
\scalebox{1}{
\begin{tabular}[b]{c@{ } c@{ }  c@{ } c@{ } c@{ } c@{ } }
    \hspace{-3mm}
    \multirow{5}{*}{\small{HR}}
    &  \includegraphics[trim={1100 550 150 900 },clip,width=.174\textwidth,valign=t]{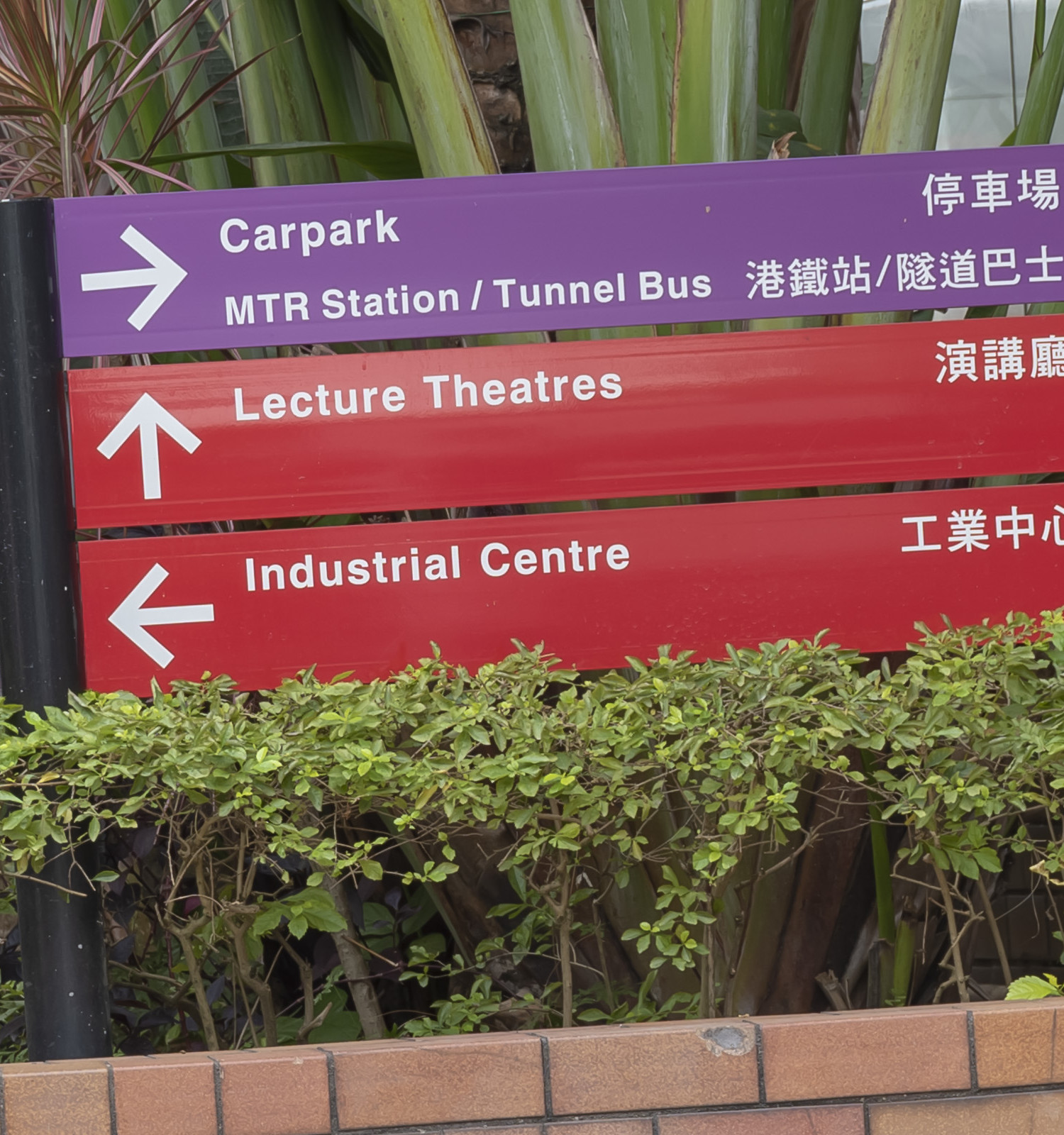}\vspace{1.2mm}
    &  \includegraphics[trim={320 180 930 870 },clip,width=.174\textwidth,valign=t]{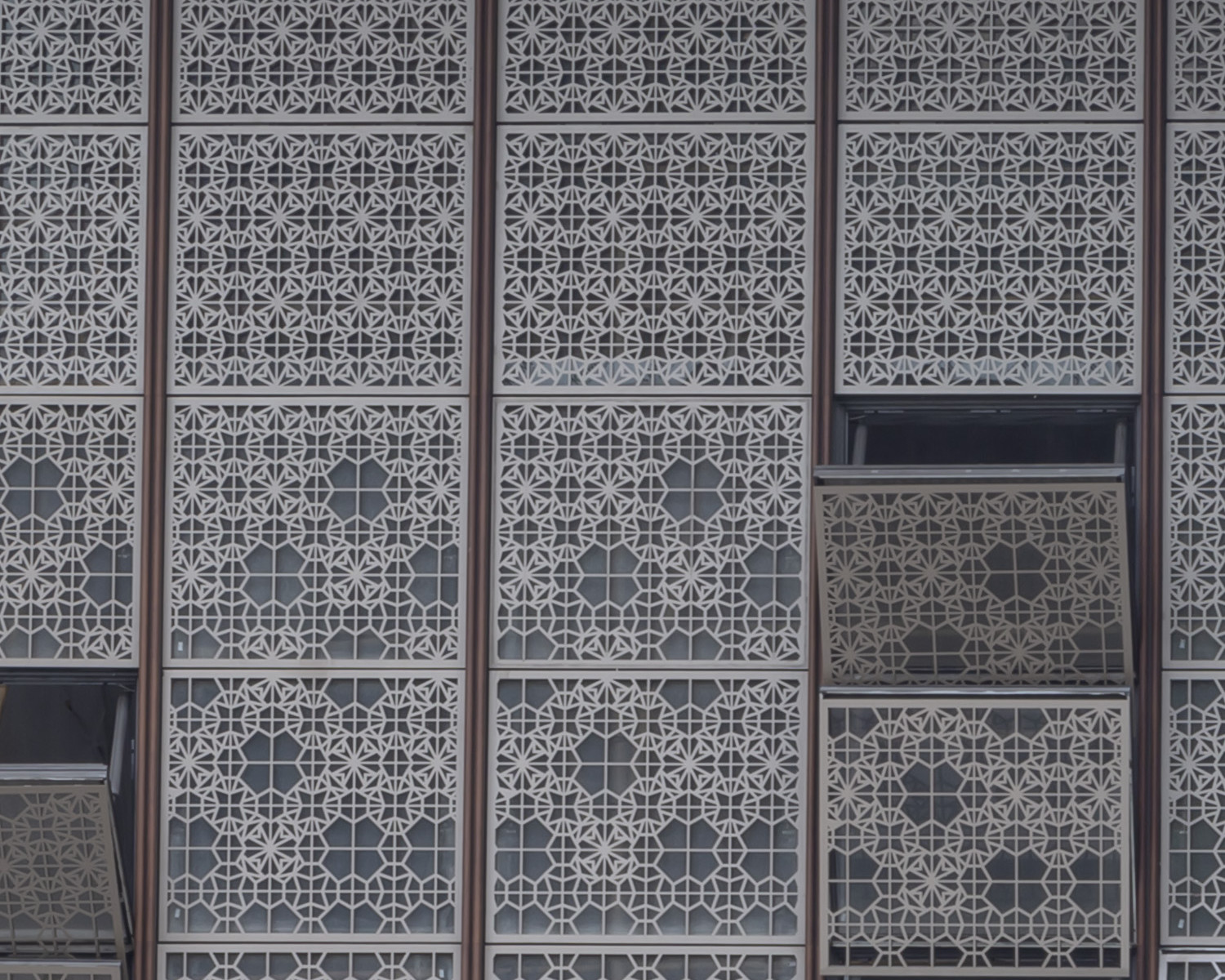}
    &  \includegraphics[trim={250 100 300 850 },clip,width=.174\textwidth,valign=t]{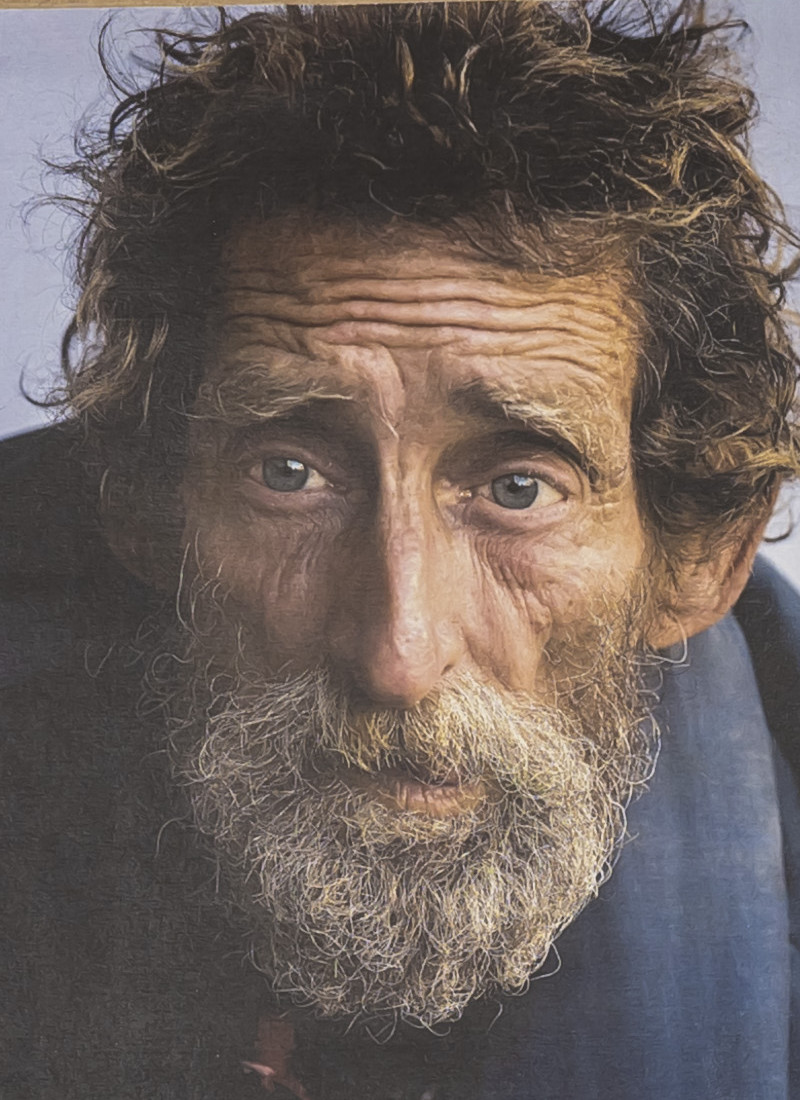}
    &  \includegraphics[trim={350 550 700 300 },clip,width=.174\textwidth,valign=t]{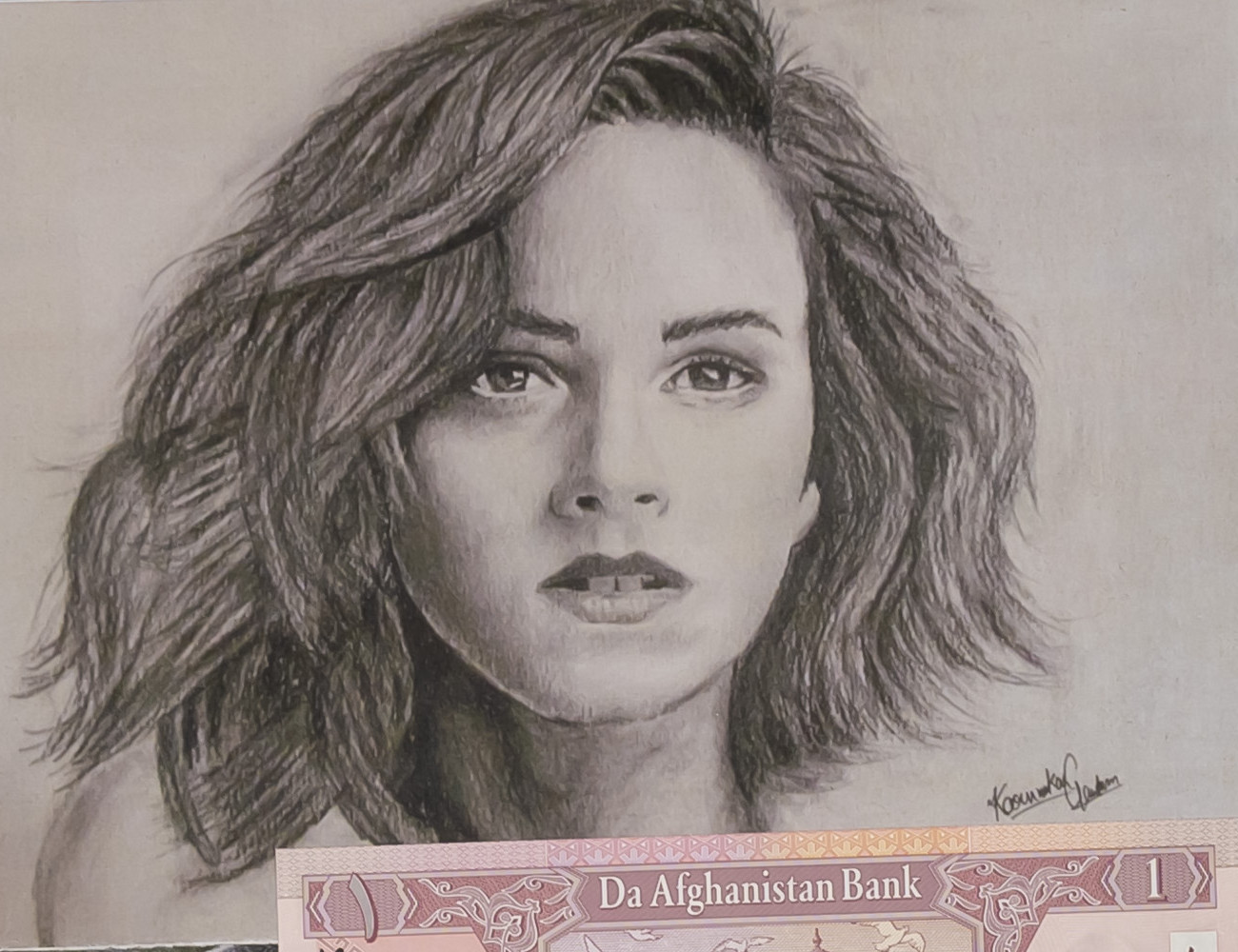}
    &  \includegraphics[trim={150 250 1700 1400 },clip,width=.174\textwidth,valign=t]{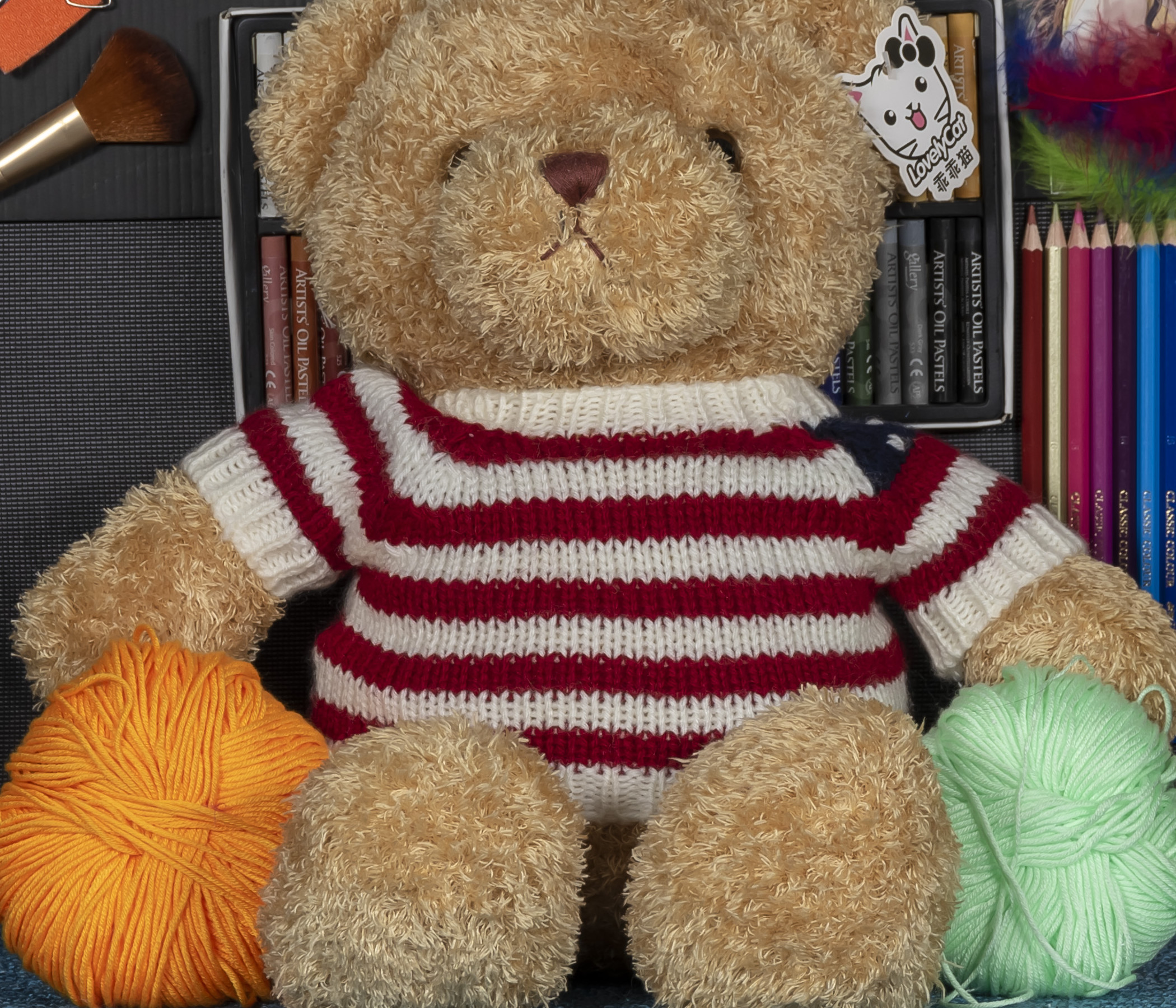}
    \\
    \hspace{-3mm}
    \multirow{6}{*}{\makecell{\small{LP-KPN} \\ \cite{RealSR}}}
    &  \includegraphics[trim={1100 550 150 900 },clip,width=.174\textwidth,valign=t]{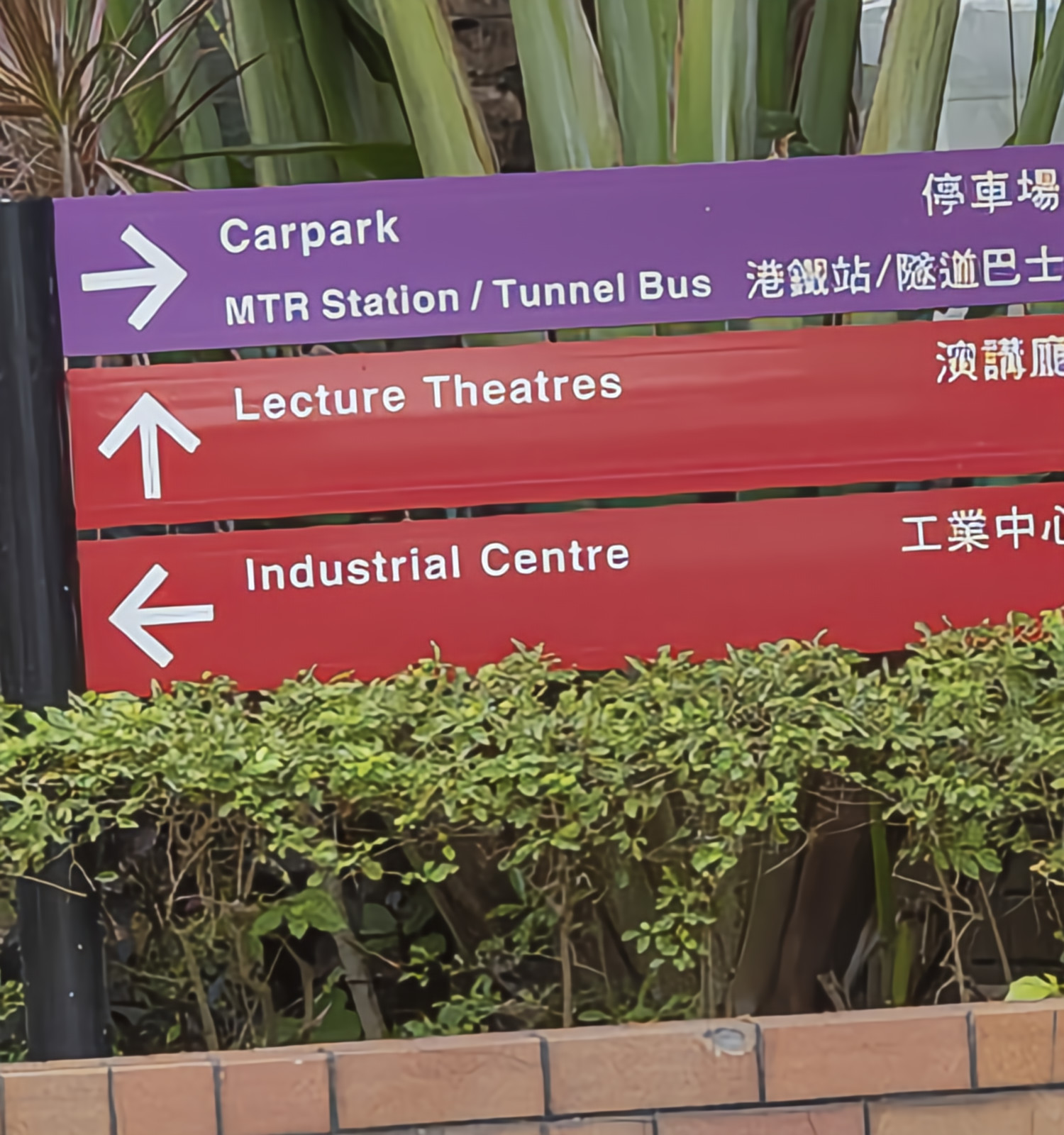}
    &  \includegraphics[trim={320 180 930 870 },clip,width=.174\textwidth,valign=t]{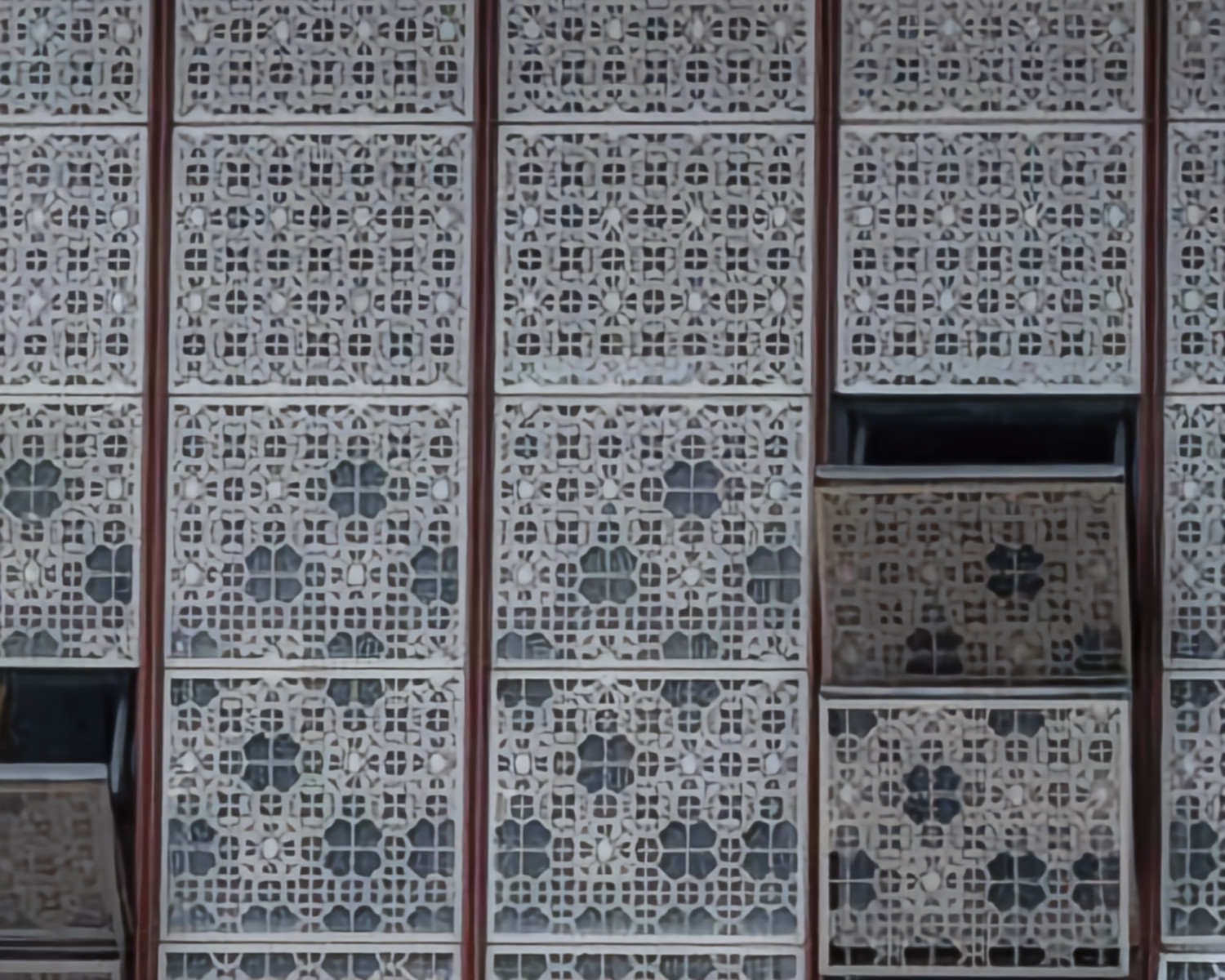}
    &  \includegraphics[trim={250 100 300 850 },clip,width=.174\textwidth,valign=t]{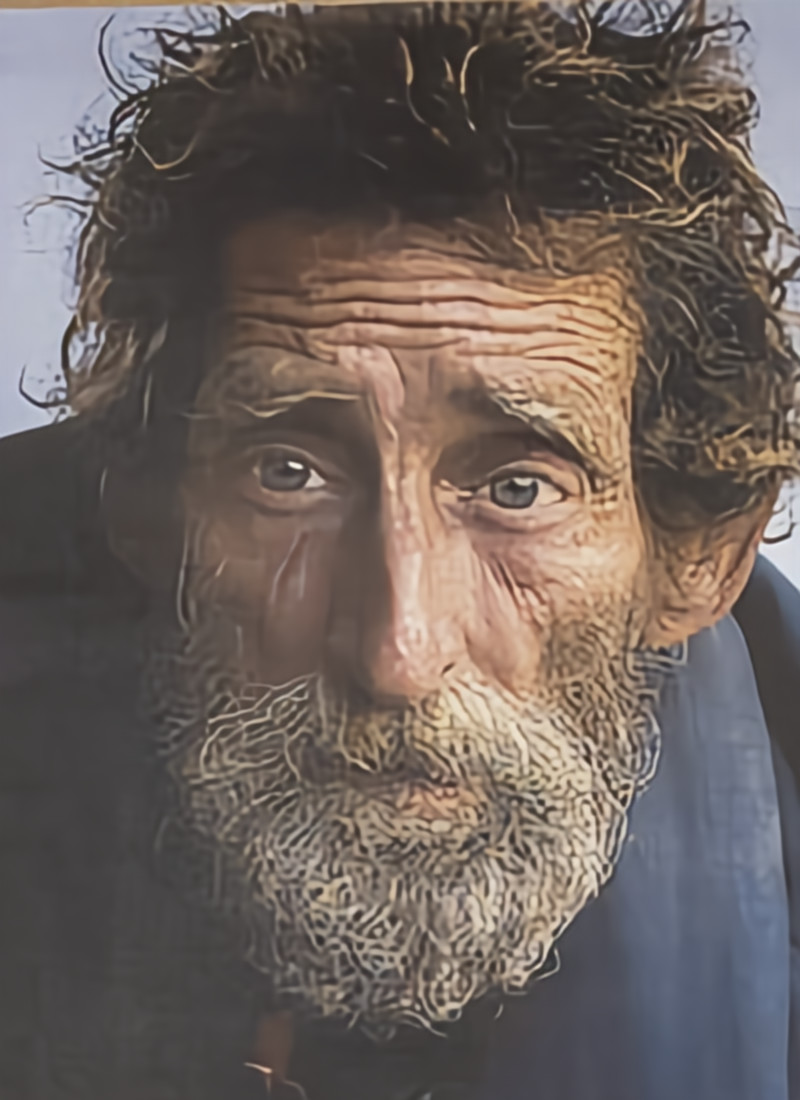}
    &  \includegraphics[trim={350 550 700 300 },clip,width=.174\textwidth,valign=t]{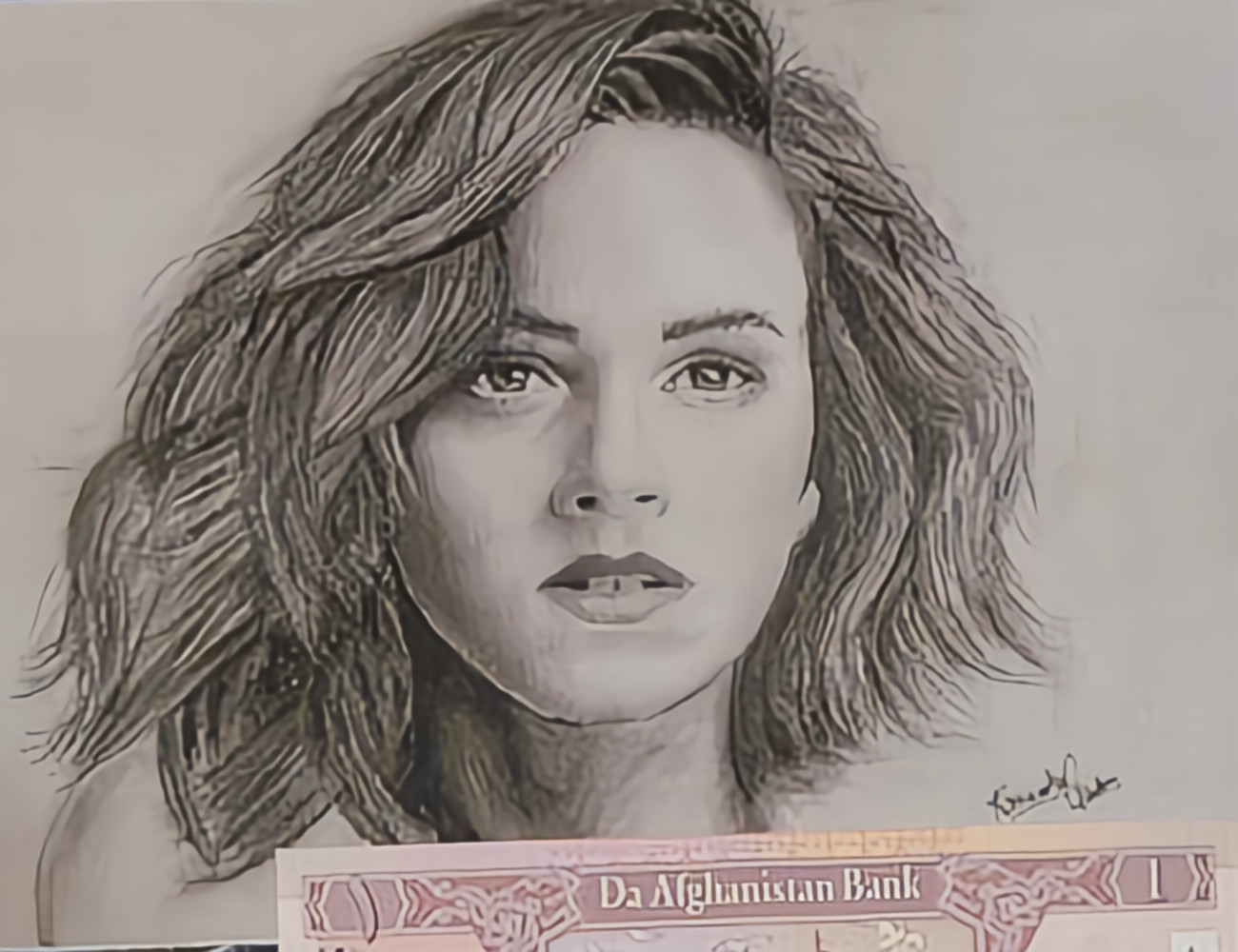}
    &  \includegraphics[trim={150 250 1700 1400 },clip,width=.174\textwidth,valign=t]{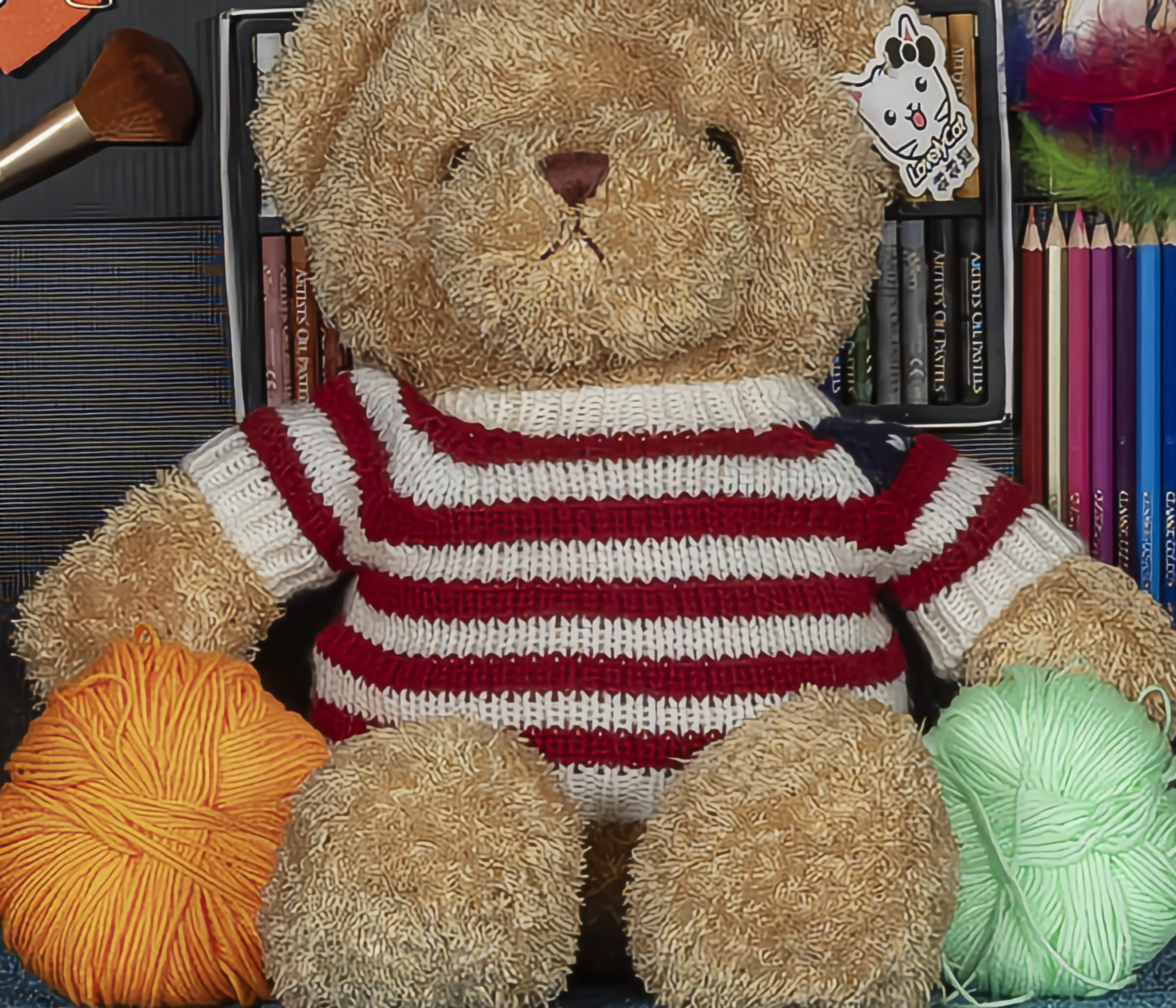}
    \vspace{0.3mm}
    \\
    \vspace{0.5mm}
    & \small{27.24 dB} & \small{21.44 dB} & \small{26.78 dB} & \small{28.55 dB} & \small{25.80 dB}
    \\
    \hspace{-3mm}
    \multirow{6}{*}{\makecell{\small{\xnet} \\ \small{(Ours)}}}
    &  \includegraphics[trim={1100 550 150 900 },clip,width=.174\textwidth,valign=t]{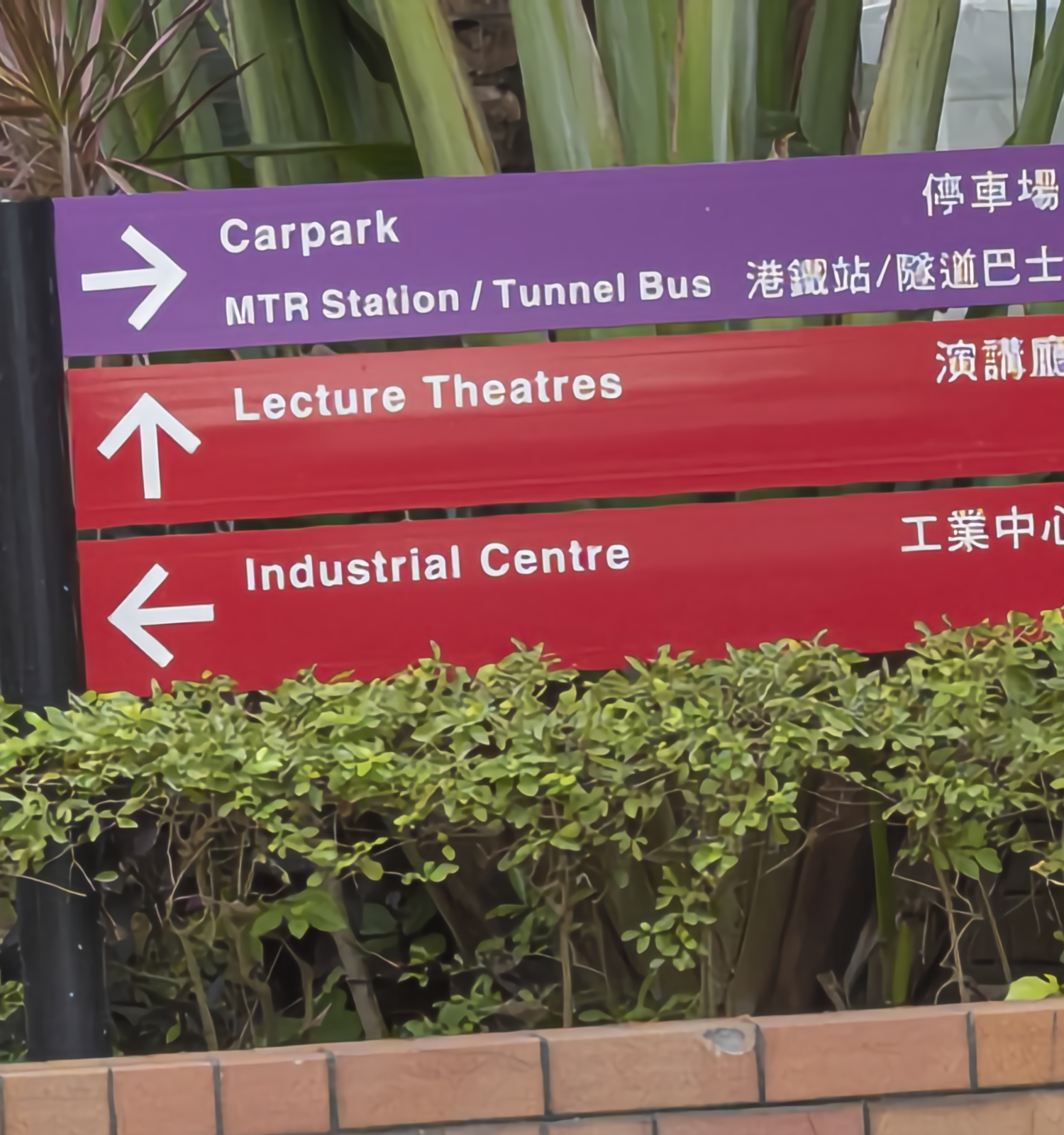}
    &  \includegraphics[trim={320 180 930 870 },clip,width=.174\textwidth,valign=t]{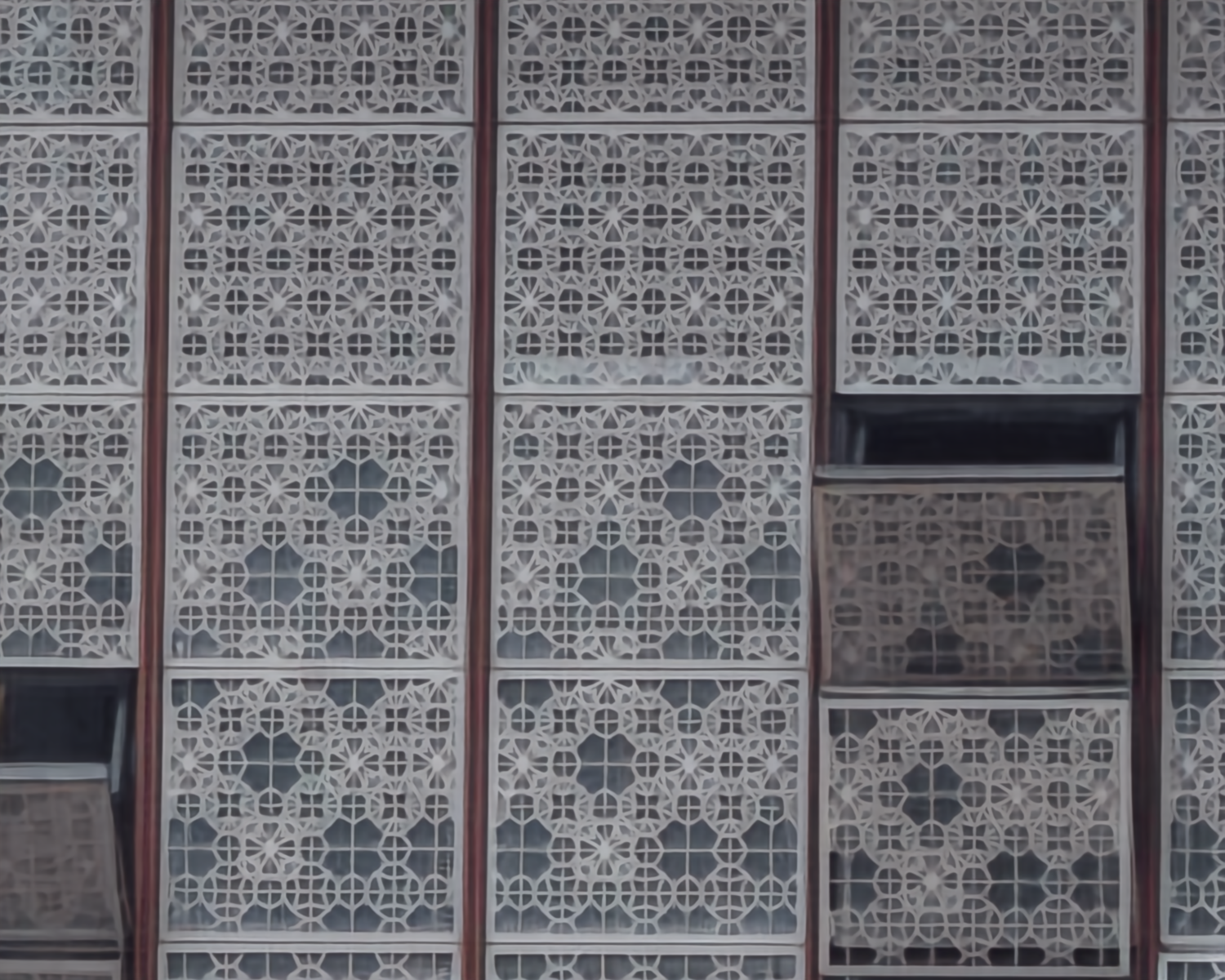}
    &  \includegraphics[trim={250 100 300 850 },clip,width=.174\textwidth,valign=t]{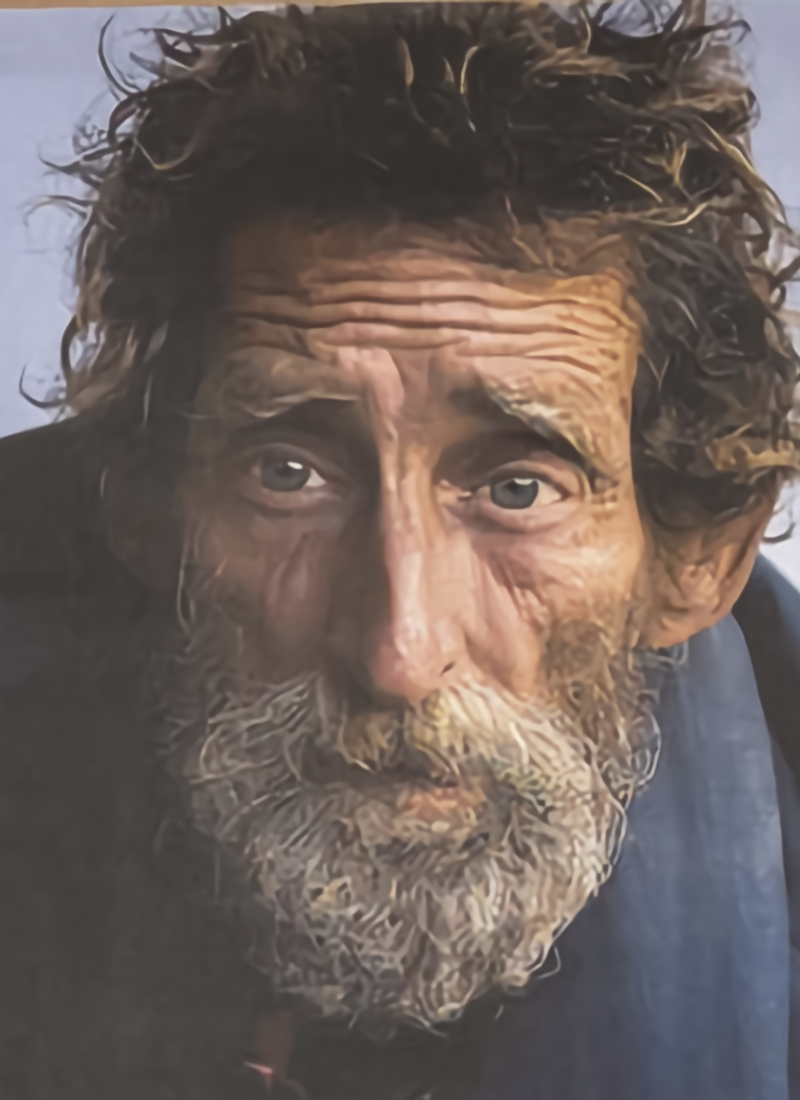}
    &  \includegraphics[trim={350 550 700 300 },clip,width=.174\textwidth,valign=t]{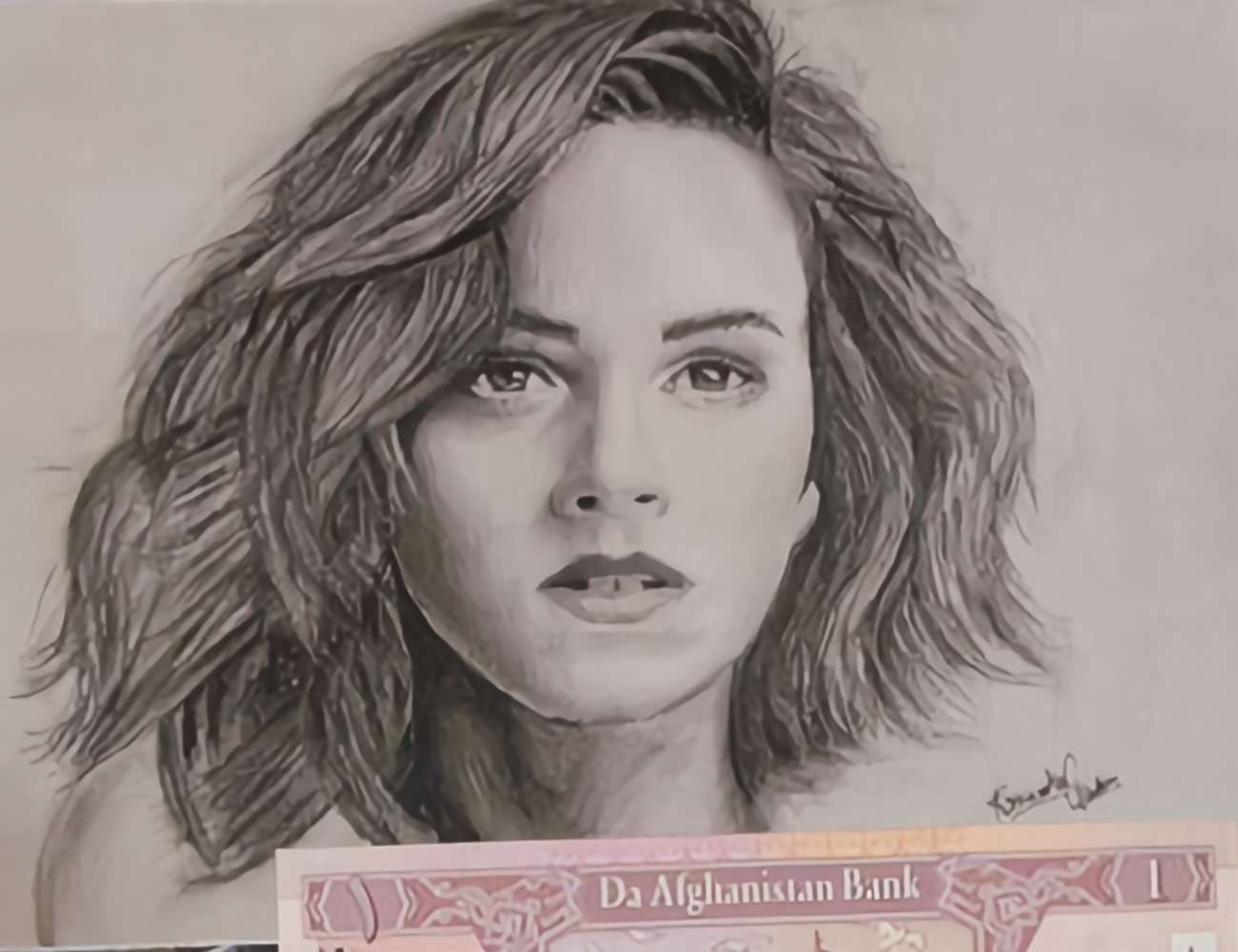}
    &  \includegraphics[trim={150 250 1700 1400 },clip,width=.174\textwidth,valign=t]{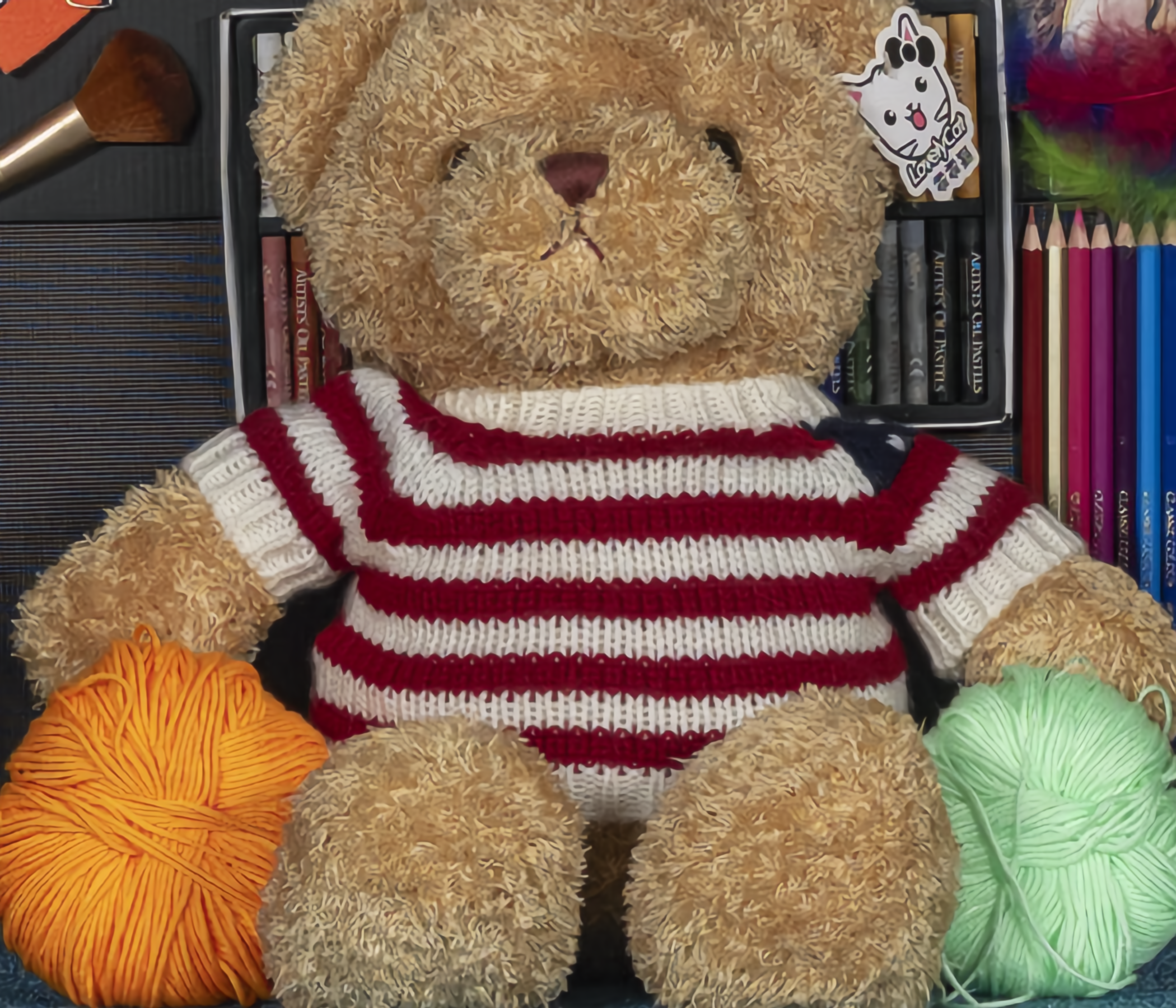}
    \vspace{0.3mm}
    \\
    & \small{\textbf{28.97 dB}} & \small{\textbf{22.88 dB}} & \small{\textbf{28.37 dB}} & \small{\textbf{31.22 dB}} & \small{\textbf{28.40 dB}}
\end{tabular}}
\end{center}
\vspace{-3mm}
\caption{Additional visual examples for $\times4$ super-resolution, comparing our \xnet against the state-of-the-art approach~\cite{RealSR}. Note that all example crops are taken from different images. 
}
\label{fig:sr crop examples}
\end{figure*}

\begin{table*}[t]
\begin{center}
\caption{Low-light image enhancement evaluation on the LoL dataset \cite{wei2018deep}. The proposed method significantly advances the state-of-the-art.}
\label{table:lol}
\setlength{\tabcolsep}{5.3pt}
\scalebox{0.99}{
\begin{tabular}{l c c c c c c c c c c c c c c }
\toprule[0.15em]
Method & BIMEF  & CRM & Dong  & LIME   & MF  & RRM  & SRIE & Retinex-Net  & MSR  & NPE & GLAD  & KinD  & KinD++ & \xnet  \\
& \cite{ying2017bio} & \cite{ying2017new} & \cite{dong2011fast} &  \cite{guo2016lime}  &  \cite{fu2016weighted} &  \cite{liu2018structure} &  \cite{fu2016weighted} &  \cite{wei2018deep} &  \cite{jobson1997multiscale} &  \cite{wang2013naturalness} &  \cite{wang2018gladnet} &  \cite{zhang2019kindling} & \cite{zhang2021beyond}  & (Ours) \\
 \midrule[0.1em]
PSNR & 13.86 & 17.20 & 16.72 & 16.76 & 18.79 & 13.88 & 11.86 & 16.77 & 13.17 & 16.97 & 19.72 & 20.87 & \underline{21.30} &\textbf{24.74} \\
SSIM & 0.577 & 0.644 & 0.582 & 0.564 & 0.642 & 0.658 & 0.498 & 0.559 & 0.479 & 0.589 & 0.703 & 0.810 & \underline{0.822} & \textbf{0.851} \\
\bottomrule[0.15em]
\end{tabular}}
\end{center}
\end{table*}

\begin{table*}[!t]
\begin{center}
\caption{Image enhancement comparisons on the MIT-Adobe FiveK dataset \cite{mit_fivek}.} 
\label{table:fivek}
\setlength{\tabcolsep}{9pt}
\scalebox{0.99}{
\begin{tabular}{l c c c c c c}
\toprule[0.15em]
Method & HDRNet \cite{Gharbi2017} & W-Box \cite{hu2018exposure} & DR \cite{park2018distort} & DPE \cite{chen2018deep}  & DeepUPE \cite{wang2019underexposed}  & \xnet (Ours) \\
 \midrule[0.1em]
PSNR & 21.96 & 18.57 & 20.97 & 22.15 & \underline{23.04} & \textbf{23.97}  \\
SSIM & 0.866 & 0.701 & 0.841 & 0.850 & \underline{0.893} & \textbf{0.931} \\
\bottomrule[0.15em]
\end{tabular}}
\end{center}
\vspace{-0.2em}
\end{table*}

Visual comparisons in Fig.~\ref{fig:sr example} show that our \xnet can effectively recover content structures . In contrast, VDSR~\cite{VDSR}, SRResNet~\cite{SRResNet} and RCAN~\cite{RCAN} reproduce results with noticeable artifacts. 
Furthermore, LP-KPN~\cite{RealSR} is not able to preserve structures (see near the right edge of the crop). 
Several more examples are provided in Fig.~\ref{fig:sr crop examples} to further compare the image reproduction quality of our method against the previous best method~\cite{RealSR}.  
It can be seen that LP-KPN~\cite{RealSR} has a tendency to over-enhance the contrast (cols. 1, 3, 4) and in turn causes loss of details near dark and high-light areas.
In contrast, the proposed \xnet successfully reconstructs structural patterns and edges (col. 2) and produces images that are natural (cols. 1, 4) and have better color reproduction (col. 5).


\subsection{Image Enhancement}

In this section, we demonstrate the effectiveness of our algorithm by evaluating it for the image enhancement task. 
We report PSNR/SSIM values of our method and several other techniques in Table~\ref{table:lol} and Table~\ref{table:fivek} for the LoL~\cite{wei2018deep} and MIT-Adobe FiveK~\cite{mit_fivek} datasets, respectively. 
It can be seen that our \xnet achieves significant improvements over previous approaches. 
Notably, when compared to the recent best methods, \xnet obtains $3.44$~dB performance gain over KinD++~\cite{zhang2021beyond} on the LoL dataset and $0.93$~dB improvement over DeepUPE\footnote{Note that the quantitative results reported in~\cite{wang2019underexposed} are incorrect. The correct scores are later released by the original authors~\href{https://drive.google.com/file/d/1fJ7MQfm6NuCMtfQzLM0Y6LNU9XyQb6Ho/view}{[link]}.} \cite{wang2019underexposed} on the Adobe-Fivek dataset.  

\begin{figure*}[!t]
  \begin{center}
  \scalebox{1}{
    \begin{tabular}{cccc}
      \includegraphics[width=0.24\textwidth]{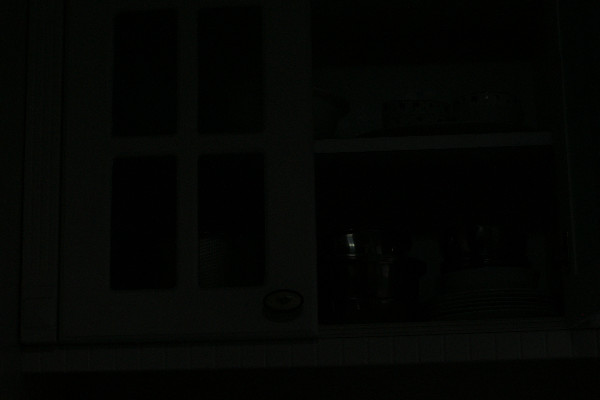}&\hspace{-3.5mm}
      \includegraphics[width=0.24\textwidth]{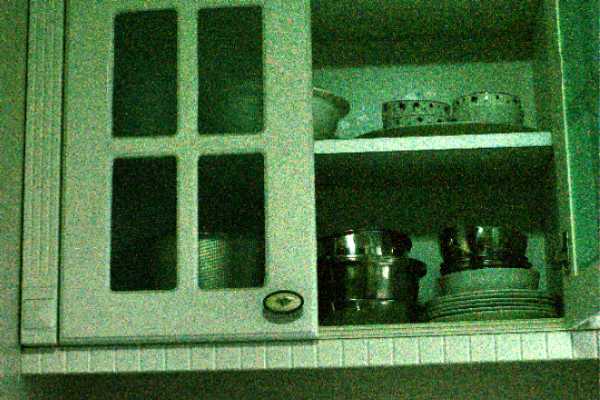}&\hspace{-3.5mm}
       \includegraphics[width=0.24\textwidth]{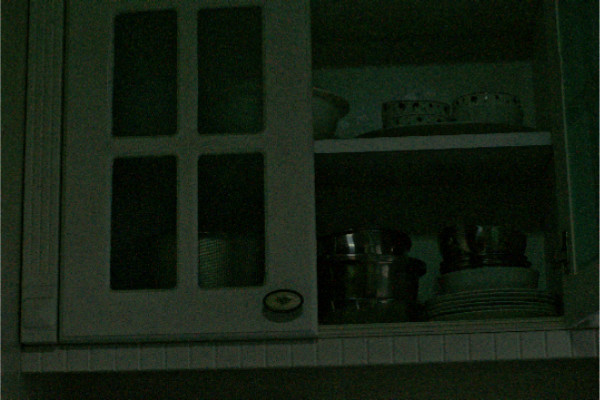}&\hspace{-3.5mm}
       \includegraphics[width=0.24\textwidth]{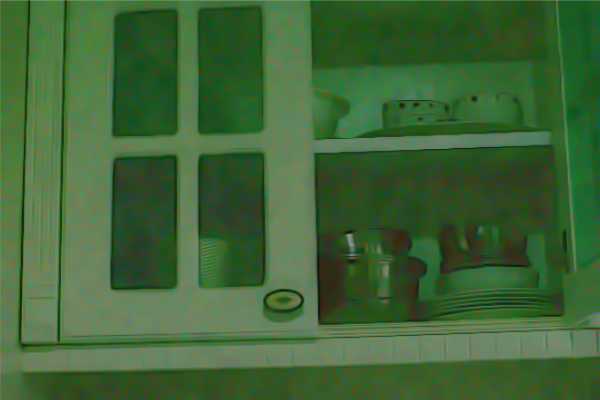}\hspace{-3.5mm}
       \vspace{0.3mm}
       \\
       \vspace{0.5mm}
      \small~Input image & \small~LIME \cite{guo2016lime} & \small~SRIE \cite{fu2016weighted} & \small~Retinex-Net \cite{wei2018deep} \\
      \includegraphics[width=0.24\textwidth]{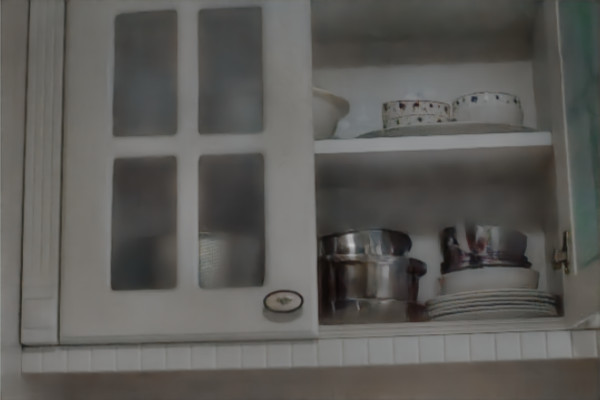}&\hspace{-3.5mm}
      \includegraphics[width=0.24\textwidth]{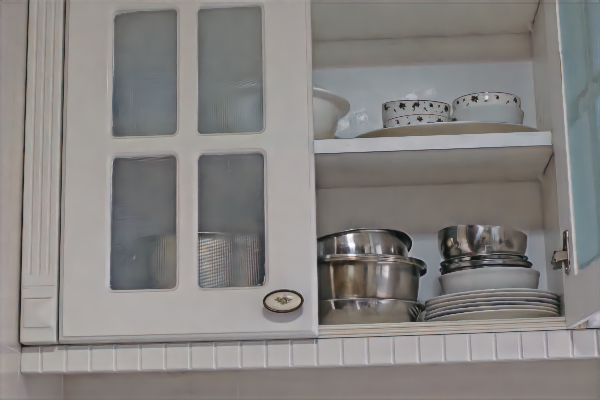}&\hspace{-3.5mm}
      \includegraphics[width=0.24\textwidth]{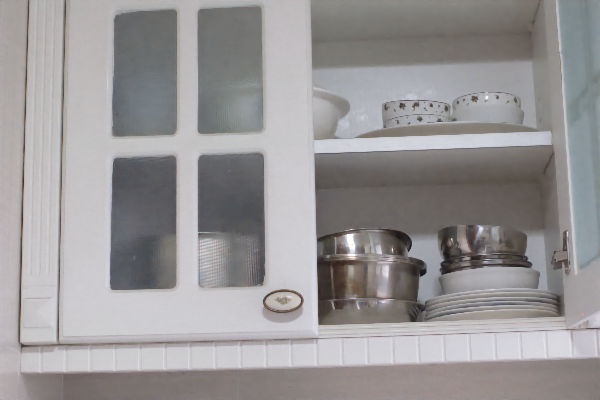}&\hspace{-3.5mm}
      \includegraphics[width=0.24\textwidth]{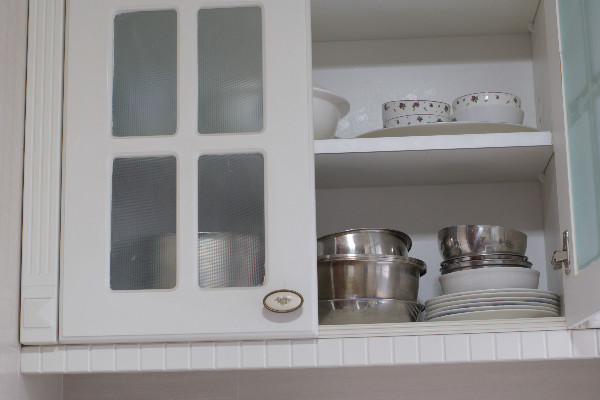}\hspace{-3.5mm}
      \vspace{0.3mm}
      \\
      \vspace{0.5mm}
       \small~KinD \cite{zhang2019kindling} & \small~KinD++ \cite{zhang2021beyond} & \small~\xnet (Ours) & \small~Ground-truth
    \end{tabular}}
  \end{center}\vspace{-3mm}
    \caption{Visual comparison of low-light enhancement approaches on the LoL dataset~\cite{wei2018deep}. The image produced by our method is visually closer to the ground-truth in terms of brightness and global contrast.}
    \label{Fig:qual_lol}
\end{figure*}

\begin{figure*}[!t]
  \begin{center}
  \scalebox{0.95}{
    \begin{tabular}{ccc}
      \includegraphics[width=0.324\textwidth]{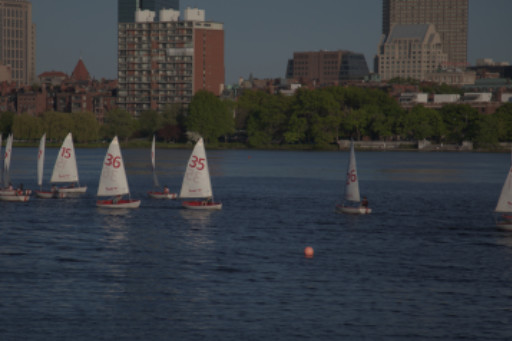}&\hspace{-1.5mm}
      \includegraphics[width=0.324\textwidth]{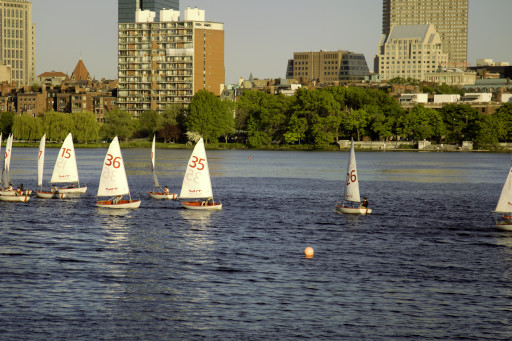}&\hspace{-1.5mm}
      \includegraphics[width=0.324\textwidth]{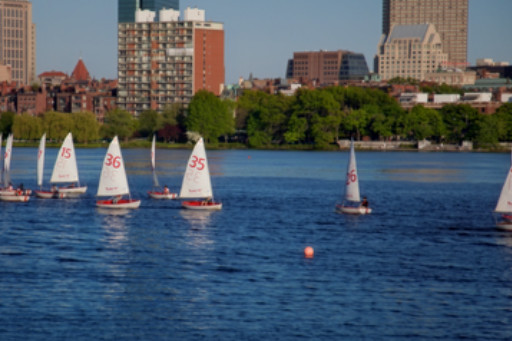}\hspace{-1.5mm}
      \vspace{0.3mm}
      \\
      \vspace{0.5mm}
      \small~Input image & \small~HDRNet~\cite{Gharbi2017} & \small~DPE~\cite{chen2018deep} \\
      \includegraphics[width=0.324\textwidth]{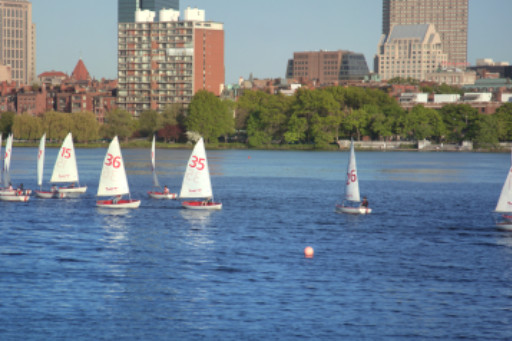}&\hspace{-1.5mm}
      \includegraphics[width=0.324\textwidth]{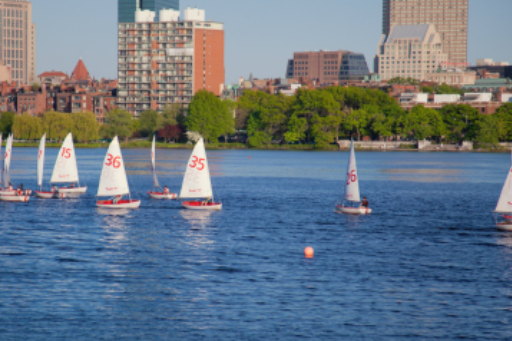}&\hspace{-1.5mm}
      \includegraphics[width=0.324\textwidth]{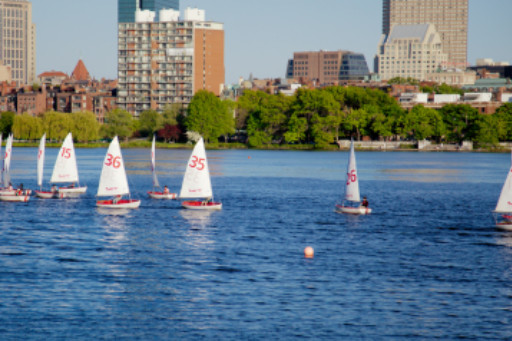}\hspace{-1.5mm}
      \vspace{0.3mm}
      \\
      \vspace{0.5mm}
       \small~DeepUPE~\cite{wei2018deep} & \small~\xnet (Ours)  & \small~Ground-truth

    \end{tabular}}
    \scalebox{0.95}{
    \begin{tabular}{ccc}
      \includegraphics[width=0.324\textwidth]{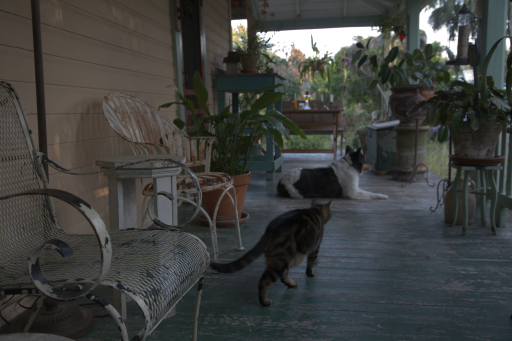}&\hspace{-1.5mm}
      \includegraphics[width=0.324\textwidth]{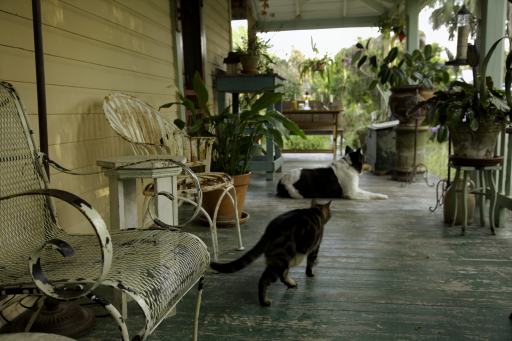}&\hspace{-1.5mm}
      \includegraphics[width=0.324\textwidth]{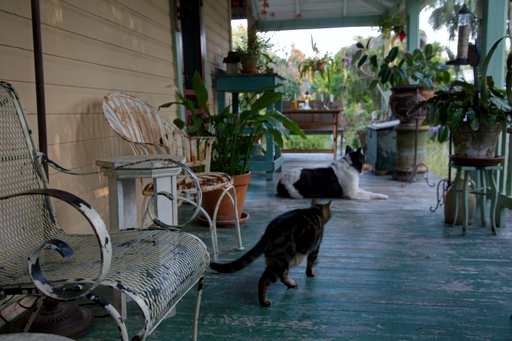}\hspace{-1.5mm}
      \vspace{0.3mm}
      \\
      \vspace{0.5mm}
      \small~Input image & \small~HDRNet~\cite{Gharbi2017} & \small~DPE~\cite{chen2018deep} \\
      \includegraphics[width=0.324\textwidth]{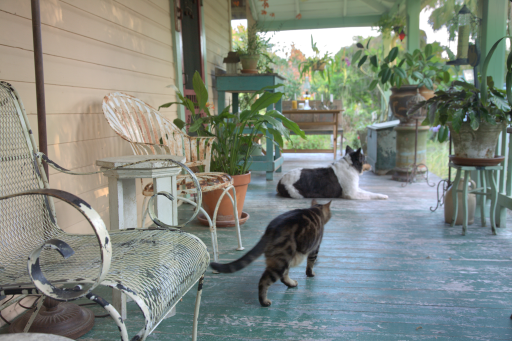}&\hspace{-1.5mm}
      \includegraphics[width=0.324\textwidth]{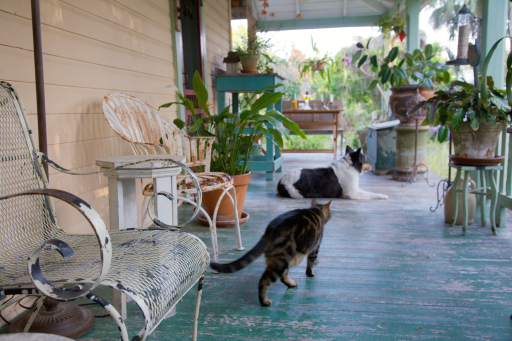}&\hspace{-1.5mm}
      \includegraphics[width=0.324\textwidth]{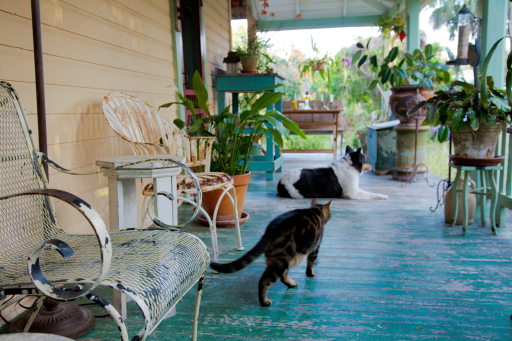}\hspace{-1.5mm}
      \vspace{0.3mm}
      \\
      \vspace{0.5mm}
       \small~DeepUPE \cite{wei2018deep} & \small~\xnet (Ours)  & \small~Ground-truth

    \end{tabular}}

  \end{center}\vspace{-3mm}
    \caption{Visual results of image enhancement on the MIT-Adobe FiveK~\cite{mit_fivek} dataset. 
    Compared to the state-of-the-art, our \xnet makes better color and contrast adjustments and produces images that appear vivid, natural and pleasant. }
    \label{Fig:qual_fivek}\vspace{-0em}
\end{figure*}

We show visual results in Fig.~\ref{Fig:qual_lol} and Fig.~\ref{Fig:qual_fivek}. 
Compared to other techniques, our method generates enhanced images that are natural and vivid in appearance and have better global and local contrast.

\subsection{Ablation Studies}
We study the impact of each of our architectural components and design choices on the final performance. 
All the ablation experiments are performed for the super-resolution task with $\times3$ scale factor. The ablation models are trained on image patches of size $128$$\times$$128$ for $10^5$ iterations.  
Table~\ref{table:ablation main} shows that removing skip connections causes the largest performance drop. Without skip connections, the network finds it difficult to converge and yields high training errors, and consequently low PSNR.
Furthermore, the information exchange among parallel convolution streams via SKFF is helpful and leads to improved performance. 
Similarly, RCB contributes positively towards the final image quality. 

Table~\ref{table:ablation rcb} shows that the proposed RCB provides favorable performance gain over the baseline Resblock from EDSR~\cite{EDSR}. Moreover, removing the transform part from RCB causes drop in accuracy. 
Table~\ref{table:ablation rcb} also shows that replacing the group convolutions with regular convolutions in RCB increases the PSNR score, but at the cost of significant increase in parameters and FLOPs. Therefore, we opt for RCB with group convolutions (g=2) as a balanced choice.  

Next, we analyze the feature aggregation strategy in Table~\ref{table:ablation aggregation}. It shows that the proposed SKFF generates favorable results compared to summation and concatenation. Note that our proposed SKFF module uses $\sim5\times$ fewer parameters than concatenation. 
Table~\ref{table:progressive} shows that the progressive learning strategy on mixed-size image patches yields PSNR similar to the model trained on large image patches (ps=224), but takes less time for training.  
Finally, in Table~\ref{table: ablation MRB} we study how the number of convolutional streams and columns (RCB blocks) of MRB affect the image restoration quality. 
We note that increasing the number of streams provides significant improvements, thereby justifying the importance of multi-scale features processing. Moreover, increasing the number of columns yields better scores, thus indicating the significance of information exchange among parallel streams for feature consolidation.   

\begin{table}[!t]
\begin{center}
\caption{Impact of individual components of MRB.}
\label{table:ablation main}
\setlength{\tabcolsep}{8pt}
\scalebox{0.99}{
\begin{tabular}{l | c c c c c}
\toprule[0.15em]
Skip connections   &      & \ch  &\ch   & \ch  &    \ch  \\ 
RCB                & \ch  &      &\ch   &      &    \ch  \\  
SKFF intermediate  & \ch  & \ch  &      &      &    \ch \\
SKFF final         & \ch  & \ch  &\ch   &\ch   &    \ch \\
\midrule[0.1em]
PSNR (in dB)   & 28.21 & 30.79 & 30.85 & 30.68 & \textbf{30.97} \\  \bottomrule[0.15em]
\end{tabular}}
\end{center}
\end{table}

\begin{table}[!t]
\begin{center}
\caption{Effect of individual components of RCB. Resblock from EDSR~\cite{EDSR} is taken as baseline. FLOPs are calculated on an image of size $256$${\times}$$256$. `g' represents the number of groups in the group convolutions.}
\label{table:ablation rcb}
\setlength{\tabcolsep}{8pt}
\scalebox{0.99}{
\begin{tabular}{l | c c c}
\toprule[0.15em]
& PSNR & Params~(M) & FLOPs~(B) \\
\midrule[0.1em]
Baseline~\cite{EDSR}, g=2   &  30.84 & 5.0 & 139.5\\ 
+ RCB, g=2   &   30.97 & 5.9 & 139.8 \\ 
RCB w/o transform, g=2   &  30.92  & 5.0 & 139.7\\ 
RCB, g=1   &  31.05  & 9.7 & 253.2\\ 
\bottomrule[0.15em]
\end{tabular}}
\end{center}
\end{table}

\begin{table}[!t]
\begin{center}
\caption{Feature aggregation. Our SKFF uses $\sim5\times$ fewer parameters than `Concat', but generates better results. }
\label{table:ablation aggregation}
\setlength{\tabcolsep}{12.5pt}
\scalebox{0.99}{
\begin{tabular}{l | c c c }
\toprule[0.15em]
    & Sum     & Concat  & SKFF  \\ 
\midrule[0.1em]
PSNR (in dB) & 30.76 & 30.83  & 30.97 \\
Parameters   & 0     & 8,192 & 1,536  \\ 
\bottomrule[0.15em]
\end{tabular}}
\end{center}
\end{table}

\begin{table}[!t]
\begin{center}
\caption{Effect of progressive learning. For progressive training, we gradually increase image patch size from $128$${\times}$$128$ to $224$${\times}$$224$. }
\label{table:progressive}
\setlength{\tabcolsep}{7pt}
\scalebox{0.99}{
\begin{tabular}{l | c c c c c  }
\toprule[0.15em]
Patch size    & 128     & 144  & 192 & 224 & Progressive \\ 
\midrule[0.1em]
PSNR (in dB) & 30.97 & 30.99 & 31.02 & 31.08 & 31.06 \\
Train time (h) & 14 & 17 & 25 & 33 & 22  \\
\bottomrule[0.15em]
\end{tabular}}
\end{center}
\end{table}

\begin{table}[!t]
\begin{center}
\caption{Ablation study on different layouts of MRB. \emph{Rows} denote the number of parallel resolution streams, and \emph{Cols} represent the number of columns containing RCBs. }
\label{table: ablation MRB}
\setlength{\tabcolsep}{10pt}
\scalebox{0.99}{
\begin{tabular}{l | c c c  }
\toprule[0.15em]
PSNR & Cols = 1 & Cols = 2 & Cols = 3  \\
\midrule[0.1em]
Rows = 1 &   30.01   &  30.29  &   30.47 \\
Rows = 2 &  30.65 &   30.79  &  30.85  \\
Rows = 3 &  30.73  & 30.97 &  31.03  \\
\bottomrule[0.15em]
\end{tabular}}
\end{center}\vspace{-2.5em}
\end{table}

\section{Concluding Remarks}

Conventional image restoration and enhancement pipelines either stick to the full resolution features along the network hierarchy or use an encoder-decoder architecture. The first approach helps retain precise spatial details, while the latter one provides better contextualized representations. However, these methods can satisfy only one of the above two requirements, although real-world image restoration tasks demand a combination of both conditioned on the given input sample. In this work, we propose a novel architecture whose main branch is dedicated to full-resolution processing and the complementary set of parallel branches provides better contextualized features. We propose novel mechanisms to learn relationships between features within each branch as well as across multi-scale branches. Our feature fusion strategy ensures that the receptive field can be dynamically adapted without sacrificing the original feature details. Consistent achievement of state-of-the-art results on six datasets for four image restoration and enhancement tasks corroborates the effectiveness of our approach.

\section*{Acknowledgements} 
Ming-Hsuan Yang is supported by NSF CAREER grant 1149783. Ling Shao is is partially supported by the National Natural Science Foundation of China (grant no. 61929104). Munawar Hayat is supported by the ARC DECRA Fellowship DE200101100.

\ifCLASSOPTIONcaptionsoff
  \newpage
\fi

\bibliographystyle{unsrt}
\bibliography{bib}

\begin{IEEEbiography}[{\includegraphics[width=1in,height=1.25in,clip,keepaspectratio]{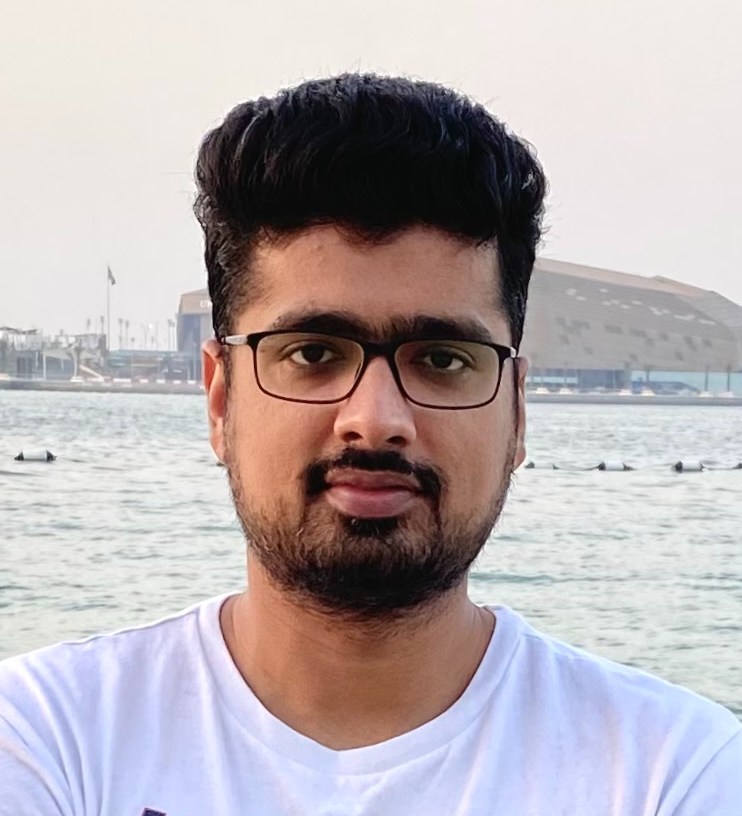}}]{Syed Waqas Zamir} 
 received the Ph.D. degree from University Pompeu Fabra, Spain, in 2017. He is a Research Scientist at Inception Institute of Artificial Intelligence in UAE. His research interests include low-level computer vision, computational imaging, image and video processing, color vision and image restoration and enhancement.
\end{IEEEbiography}

\begin{IEEEbiography}[{\includegraphics[width=1in,height=1.25in,clip,keepaspectratio]{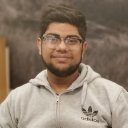}}]{Aditya Arora} 
is a Research Engineer at Inception Institute of Artificial Intelligence in UAE. His research interests include image and video processing, computational photography and low-level vision.
\end{IEEEbiography}

\begin{IEEEbiography}[{\includegraphics[width=1in,height=1.25in,clip,keepaspectratio]{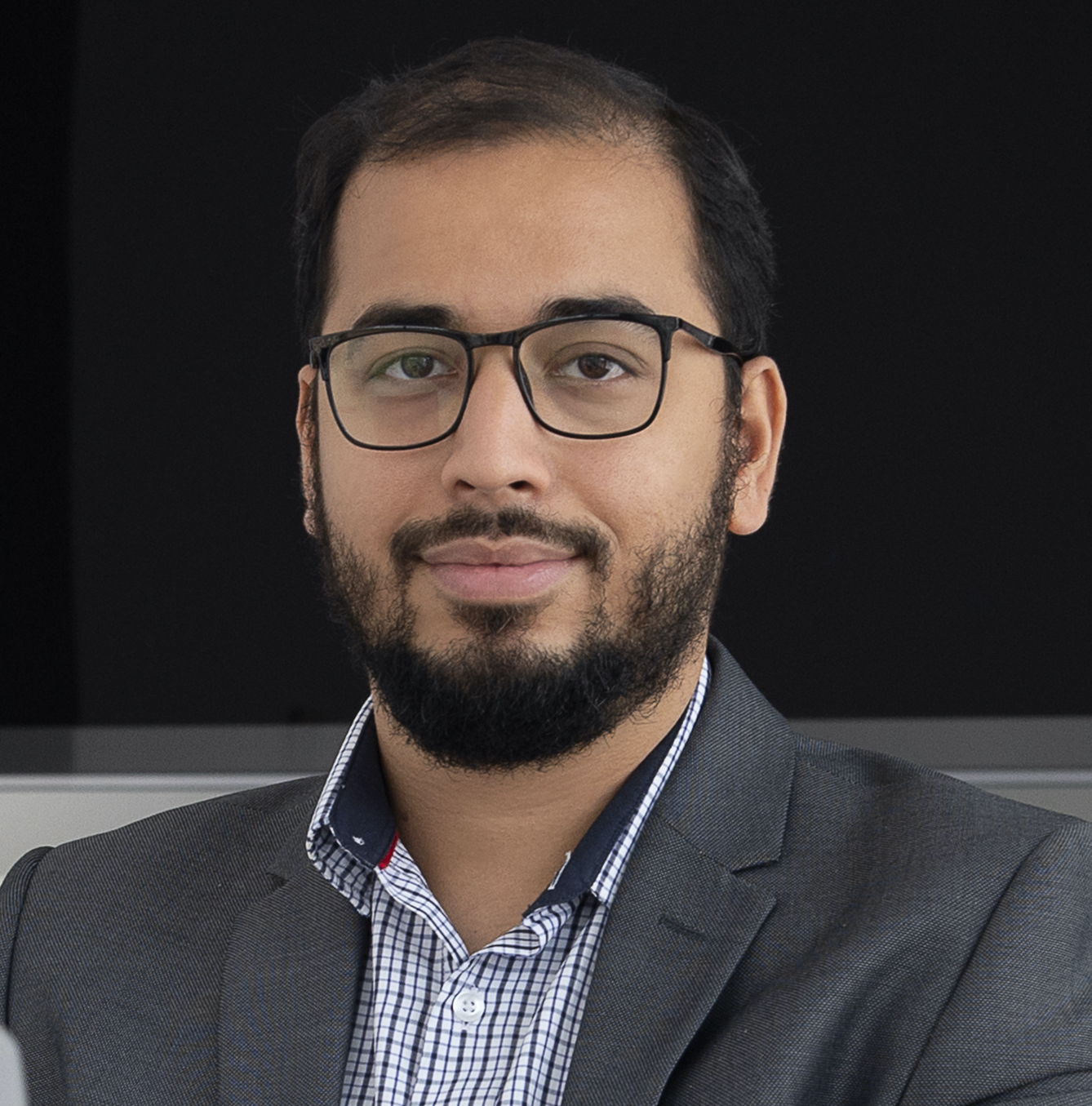}}]{Salman Khan}
is an Assistant Professor at MBZ University of Artificial Intelligence. He has been an Adjunct faculty member with Australian National University since 2016. 
He has been awarded the outstanding reviewer award at CVPR multiple times, won the best paper award at 9th ICPRAM 2020, and 2nd prize in the NTIRE Image Enhancement Competition at CVPR 2019. He served as a program committee member for several premier conferences including CVPR, ICCV, ICLR, ECCV and NeurIPS. He received his Ph.D. degree from the University of Western Australia in 2016. His thesis received an honorable mention on the Dean’s List Award. 
His research interests include computer vision and machine learning.
\end{IEEEbiography}

\begin{IEEEbiography}[{\includegraphics[width=1in,height=1.25in,clip,keepaspectratio]{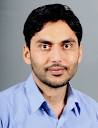}}]{Munawar Hayat}
 received his PhD from The University of Western Australia (UWA). His PhD thesis received multiple awards, including the Deans List Honorable Mention Award and the Robert Street Prize. After his PhD, he joined IBM Research as a postdoc and then moved to the University of Canberra as an Assistant Professor. He is currently a Senior Scientist at Inception Institute of Artificial Intelligence, UAE. Munawar was granted two US patents, and has published over 30 papers at leading venues in his field, including TPAMI, IJCV, CVPR, ECCV and ICCV. His research interests are in computer vision and machine/deep learning.
\end{IEEEbiography}

\begin{IEEEbiography}[{\includegraphics[width=1in,height=1.25in,clip,keepaspectratio]{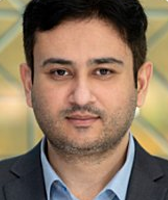}}]{Fahad Khan}
is a faculty member at MBZUAI, United Arab Emirates and Linkoping University, Sweden. From 2018 to 2020 he worked as a Lead Scientist at the Inception Institute of Artificial Intelligence (IIAI), Abu Dhabi, United Arab Emirates. He received the M.Sc. degree in Intelligent Systems Design from Chalmers University of Technology, Sweden and a Ph.D. degree in Computer Vision from Autonomous University of
Barcelona, Spain. He has achieved top ranks on various international challenges (Visual Object Tracking VOT: 1st 2014 and 2018, 2nd 2015, 1st 2016; VOT-TIR: 1st 2015 and 2016; OpenCV Tracking: 1st 2015; 1st PASCAL VOC 2010). His research interests include a wide range of topics within computer vision and machine learning, such as object recognition, object detection, action recognition and visual tracking. He has published articles in high-impact computer vision journals and conferences in these areas. He serves as a regular program committee member for leading computer vision conferences such as CVPR, ICCV, and ECCV.
\end{IEEEbiography}

\begin{IEEEbiography}[{\includegraphics[width=1in,height=1.25in,clip,keepaspectratio]{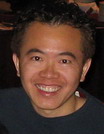}}]{Ming-Hsuan Yang} is affiliated with Google, UC Merced, and Yonsei University.  
Yang serves as a program co-chair of IEEE International Conference on Computer Vision (ICCV) in 2019, program co-chair of Asian Conference on Computer Vision (ACCV) in 2014, and general co-chair of ACCV 2016. Yang served as an associate editor of the IEEE Transactions on Pattern Analysis and Machine Intelligence, and is an associate editor of the International Journal of Computer Vision, Image and Vision Computing and Journal of Artificial Intelligence Research. He received the NSF CAREER award and Google Faculty Award. He is a Fellow of the IEEE.
\end{IEEEbiography}

\begin{IEEEbiography}[{\includegraphics[width=1in,height=1.25in,clip,keepaspectratio]{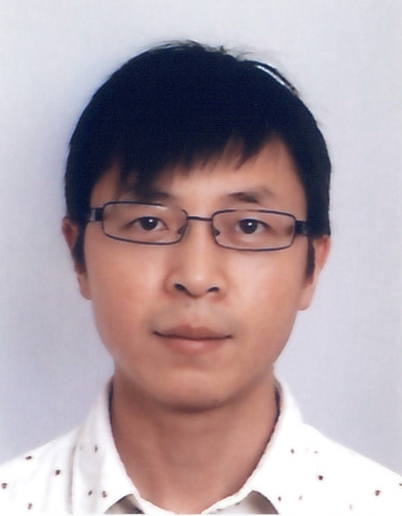}}]{Ling shao} 
is the Chief Scientist of Terminus Group  and the President of Terminus International. He was the founding CEO and Chief Scientist of the Inception Institute of Artificial Intelligence, Abu Dhabi, UAE. His research interests include computer vision, deep learning, medical imaging and vision and language. He is a fellow of the IEEE, the IAPR, the BCS and the IET.
\end{IEEEbiography}








\end{document}